\documentclass[12pt,letterpaper]{article}
\usepackage{floatrow}
\newfloatcommand{capbtabbox}{table}[][\FBwidth]

\newcommand{\footremember}[2]{%
  \footnote{#2}
  \newcounter{#1}
  \setcounter{#1}{\value{footnote}}%
}
\newcommand{\footrecall}[1]{%
  \footnotemark[\value{#1}]%
} 

\usepackage{pslatex}
\usepackage{apacite}

\usepackage{kpfonts}
\usepackage{stackengine,scalerel}
\usepackage{calc}
\newlength\shlength
\newcommand\xshlongvec[2][0]{\ThisStyle{\setlength\shlength{#1\LMpt}%
 \stackengine{-5.6\LMpt}{$\SavedStyle#2$}{\smash{$\kern\shlength%
  \stackengine{\dimexpr 1.3pt+6.25\LMpt}{$\SavedStyle\mathchar"017E$}%
   {\rule{\widthof{$\SavedStyle#2$}}{\dimexpr.1pt+.5\LMpt}\kern.4\LMpt}{O}{r}{F}{F}{L}\kern-\shlength$}}%
   {O}{c}{F}{T}{S}}}
   
\usepackage{array} 

\usepackage{caption}
\usepackage{graphicx}
\usepackage{amssymb}
\usepackage{lineno}

\title{L\'{e}vy Flights of the Collective Imagination}

\author{
William H. W. Thompson\footremember{sjc}{St.\ John's College, Santa Fe, NM 87505, USA} \and
Zachary Wojtowicz\footremember{sds}{Department of Social and Decision Sciences, Dietrich College, Carnegie Mellon University, 5000 Forbes Avenue, Pittsburgh, PA 15213, USA} \and
Simon DeDeo\footrecall{sds} \footremember{sfi}{Santa Fe Institute, 1399 Hyde Park Road, Santa Fe, NM 87501, USA; {\tt sdedeo@andrew.cmu.edu}.}
}
\date{}

\begin{document}

\maketitle

\begin{abstract}
\noindent We present a structured random-walk model that captures key aspects of how people communicate in groups. Our model takes the form of a correlated L\'{e}vy flight that quantifies the balance between focused discussion of an idea and long-distance leaps in semantic space. We apply our model to three cases of increasing structural complexity: philosophical texts by Aristotle, Hume, and Kant; four days of parliamentary debate during the French Revolution; and branching comment trees on the discussion website Reddit. In the philosophical and parliamentary cases, the model parameters that describe this balance converge under coarse-graining to limit regions that demonstrate the emergence of large-scale structure, a result which is robust to translation between languages. Meanwhile, we find that the political forum we consider on Reddit exhibits a debate-like pattern, while communities dedicated to the discussion of science and news show much less temporal order, and may make use of the emergent, tree-like topology of comment replies to structure their epistemic explorations. Our model allows us to quantify the ways in which social technologies such as parliamentary procedures and online commenting systems shape the joint exploration of ideas.

\vspace{2mm}
\end{abstract}


\noindent A characteristic feature of our species is the capacity to share and thereby collaboratively process information~\cite{Tomasello}. Overlaid on this basic capacity is an ability to create novel institutions to manage these interactions: we not only think together, but have the capacity to invent new ways to do it~\cite{searle2010making}.

Cultures are in part defined by these institutions~\cite{chwe2013rational}: we might speak in public or before an assembly; participate in a seminar or debate an opponent; publish books and blog posts, or circulate letters. While some constraints on potential institutions---such as the cognitive constraints that govern how quickly we process information---always apply, the modern era has increased our options for structuring social communication at an ever-increasing pace. The ways we can share and argue about our ideas only seem to be limited by our imaginations and the speed at which we write new software.

Laboratory experiments show that the structure governing a group's interaction determines how well they can solve problems \cite{goldstone2013learning}. A basic paradigm for these studies is the balance between exploration (seeking out new ideas) and exploitation (refining an idea currently in play~\cite{pirolli2007information}). This is an epistemic version of the problem that animals face when foraging for food~\cite{nassar2010approximately}. Here the resource is information itself, and individuals must decide when an idea is worth further investigation or has been played out and should be dropped for the time being \cite{sang_todd_goldstone_hills_2018}. When group-level structures are present, they can alter the patterns of interaction among individuals and lead to behavior that differs from what happens when they act in isolation. For example, as some problems become more complex, individuals find better solutions when they are constrained to interact with just a subset of the others in the group \cite{wisdom2013social}. 

Work in the laboratory allows us to prescribe and vary precisely-calibrated conditions to test for these effects. This comes at the cost of simplifying the kinds of problems that groups in the real world are actually concerned with solving, and the scales at which we can work. Techniques from data science now allow us to complement experimental work with studies of complex, real-world social problem solving. The exploration/exploitation paradigm has been applied, for example, to the case of individual-level foraging in a larger culture~\cite{murdock2017exploration}, large-scale online systems~\cite{lorince2016music}, and the process of scientific innovation~\cite{youn2015invention}. 

Groups that develop ideas together face additional challenges arising from cognitive and social constraints on the transmission and comprehension of information. These limitations bind at all scales, forming a hierarchy of problems that must be solved simultaneously by both individual minds and the institutions governing their communication~\cite{jackendoff2002foundations}. At the sentence level, syntax promotes the efficiency of information exchange through compressed serialization, appropriate redundancy, and minimal ambiguity \cite{chomsky2014aspects, wedel4, wedel1, wedel2, wedel3, wedel5}. At the level of the paragraph, authors present logical arguments, make emotional appeals, and establish authority~\cite{aristotle}, and rely on rhetorical tropes such as parallelism, antithesis, and chiasmus~\cite{lanham2012handlist}. Social conventions operate at larger scales yet, facilitating language production and structuring social activity \cite{jackendoff2007language}. For example, it is impossible for a group to attend to two speakers simultaneously, which leads to the creation of parliamentary rules to establish procedures for transferring speaking rights. Speakers themselves are expected to take consistent positions, argue a single point, and provide additional nonverbal signals that aid comprehension. Such rules and norms are critical to the successful use of language, even when they are not made explicit \cite{grice1975logic}.

Little work has been done on the comparative analysis of how real-world groups manage the process of collectively searching the space of ideas. This limits our ability to assess the cognitive effects of innovations to social communication, both historical and modern. We do not understand, for example, what happens when the investigation of ideas moves from the didactic lecture to the seminar table or the parliamentary debate, and then, in the modern era, to online systems that allow for branching interactions where individuals can spawn parallel discussions when they reply to a comment. 

This paper presents and applies a simple two-parameter model for how groups explore and exploit the space of ideas. Our model uses machine learning to represent the ideas contained in a collection of texts as points in a structured semantic space. This allows us to apply our model to any system where the evolution of a discussion amounts to a sequence of speech acts. We can then compare different systems by comparing the inferred parameters of the model.

Having presented this model, we then apply it to real-world discussions. Our data come from three sources, each corresponding to a distinct structure for organizing idea creation and transmission. Our first set is a collection of philosophical texts. Texts like these can be considered ``one-to-many'' discussions: the author attempts to structure the presentation of her ideas for a larger community of readers. That community has no ability to react to and thereby affect the structure of an argument as it unfolds. This set provides a baseline against which to compare social discussions, where multiple agents dynamically co-construct the investigation.

Our second set is a collection of debates from the parliament convened during the French Revolution that began in 1789. In this case, the course of debate is co-constructed by multiple speakers. Speakers may disagree about how to investigate the ideas in play, leading to more complicated ways of navigating the exploration-exploitation trade off. At the same time, debates, like the philosophical texts, have a fixed sequence; speeches occur one after another, governed by a process for determining order that all members have agreed upon. All attentive listeners will hear Speech A followed by Speech B, although the speeches themselves may wander between topics. In this sense, the debate itself is linear.

Our final set is a collection of discussions from the website Reddit. As in the case of parliamentary debate, the speech acts in each discussion are produced by a large number of agents who may have different strategies for navigating the space of possible ideas in play. In contrast to both the philosophical texts and the parliamentary debates, however, discussions can branch. Comment A, for example, can receive two replies (B and C), and each participant must choose where to allocate her attention and to which comment she replies. Comments still have a unique time-sequence (Comment B was added either before, or after Comment C), but the order in which they are encountered by participants is no longer fixed and universal. Discussions on Reddit form a branching, tree-like structure which allows a social exchange of ideas to proceed in a parallel fashion, distributed across many minds.

Our model reduces each of these datasets to a time-ordered stream of speech acts. We deliberately drop the explicit markers (\emph{e.g.}, change of speaker, in the case of debates) that structure this stream in order to detect the overall effects of the metastructure on the underlying stream. The branching conversation trees of Reddit, for example, may be able to parse and sort arriving information so as to accommodate a far less structured stream than a listener in parliament. Similarly, the salient marking of a shift of speaker in a debate may make it possible for a listener to follow more disconnected and associative paths than in a single-author text: a listener may be able to follow larger leaps if they are signposted by, say, the stepping down of a liberal speaker and his replacement by a conservative. In both cases, these structures affect the nature of the streams of information they create and then, potentially, absorb.

We expect the surprise incurred by a long jump to have a number of effects. High levels of surprise are known to draw attention~\cite{itti2009bayesian}. They are also associated with cognitive load: in a simple Bayesian model where distance is quantified by Kullback-Leibler divergence, long jumps are equivalent to coding failure, or the temporary inability to assimilate new patterns of information in an efficient fashion. Finally, while high surprise events may draw attention, they tend to do so only temporarily: such an event often fails to influence the properties of events downstream. What is very surprising is also, often, quickly forgotten~\cite{barron,elise}.

Long jumps in semantic space are not necessarily bad. They may be associated with particularly illuminating moments that draw together otherwise distinct ideas. Some arguments draw both their force and their meaning from this process. The point where a text synthesizes a number of distinct strands of argument is often a celebrated and crucial moment~\cite{murdock2017exploration}. In a debate, a speaker who manages to connect the concerns of opposing sides may carry the day.

All of these questions depend upon the size of the units in question, because different principles of organization govern a text at different scales. The development of an argument from sentence to sentence is very different, for example, from how the different parts of that argument are fit together. We view this additional degree of freedom as a valuable tool that enables the researcher to probe different layers of a social system. Thus, we examine the dynamics of the three systems as a function of the size of the unit of analysis (``chunk'') and chart the parameter estimates at each scale. We consider a range of chunk sizes in this analysis, from twenty-five words up to two hundred and fifty. Since these units are defined after the removal of stop words, they correspond to a range of lengths from a sentence or two up to multiple paragraphs. To build intuition, the abstract of this paper would roughly correspond to a unit that was 100 words in size after basic text cleaning was applied, and in our data, the overwhelming majority of chunks at this size contain at least three sentences.

Analyzing the system in terms of larger and larger chunks, or working with a more ``coarse-grained'' description, allows us to see how the cognitive task of attending and sense-making changes with scale. It may, for example, be the case that sentences lock together in particularly determinate and predictable ways, producing lego-like meaning-units that can then fill in the larger-scale structure in an more arbitrary fashion with long leaps from one area to another. Conversely, sentences are expected to be processed more quickly than paragraphs, because their comprehension is aided by rapid sense-making system-one heuristics~\cite{sloman1996empirical}. This may mean that more rapid shifts back and forth in semantic space are possible, while the slower system-two mechanisms that operate on the level of argument construction need to provide greater guidance and structure.

\section{Methods}

For our analysis, we: (i) construct datasets from three collections of texts, (ii) semantically decompose these collections using topic modeling, and (iii) estimate our structured random-walk model on these semantic representations using Bayesian inference. We describe each of these steps in turn.


\subsection{Data and Topic Modeling}

Our first dataset is three major books each from the philosophers Aristotle, David Hume, and Immanuel Kant. For Aristotle, we use the \emph{Metaphysics}, the \emph{Nichomachean Ethics}, and the \emph{Rhetoric}; for Hume, \emph{An Enquiry Concerning Human Understanding}, \emph{A Treatise of Human Nature}, and \emph{An Enquiry Concerning the Principles of Morals}; for Kant, the \emph{Critique of Pure Reason}, the \emph{Critique of Practical Reason}, and the \emph{Critique of Judgement}. The texts are taken from the Gutenberg Project,\footnote{\url{http://gutenberg.org}} which contains public-domain translations that are then checked by an online, distributed team of proofreaders.

For our second collection, we take four separate days of debates during the French Revolution during which discussion was centered around a single topic. These four dates are chosen so that there are at least sixty chunks of size 250.

The day with the most words occurs on 21 June 1971, the day after the ``Flight to Varennes'', when the King and Queen of France attempted, in secret, to escape the country to raise an army and reassert their power. The Flight to Varennes was a crucial turning point in the revolution. It humiliated many who had urged tolerance of the aristocratic system and a shift to a British-style constitutional monarchy~\cite{tackett2015coming}. All other discussion was put aside that day as liberals, conservatives, and a soon-to-grow hard-left argued over how to respond. The other three days, 22 March, 5 May, and 18 June 1791 are more typical, and debates on those days ranged across different topics of concern such as budget deficits, the nationalization of church property, and unrest among military officers.

In these debates, we exclude ``presidential'' speech (\emph{i.e.}, short procedural statements by the organizing member), and treat the remainder as a continuous stream, dropping markers associated with a change of speaker. These reduced files are drawn from the analysis of \citeA{barron}. The original texts are in French, so to determine the extent to which the language of the source affects the derived parameters of the models below we conduct a simultaneous analysis using a (machine-authored) English-language version of both days, produced using the Google Translate interface.

For our branching case, we take fifteen ``submissions'' from the online discussion site Reddit. A submission is a branching discussion initiated by a short opening paragraph, or a URL to somewhere else on the internet. Each set includes both the initiating post of the interaction (the submission itself) and the subsequent, downstream discussion as people post comments in reply. The submissions are drawn from three different ``subreddits'', forums with distinct subject matter and social norms: r/The\_Donald, r/science, and r/news. These were chosen to span a range of different discussion goals: political argument-making, explanation-making and evaluation, and information gathering and sharing. The subreddit r/The\_Donald is devoted to the celebration of the political positions associated with the U.S.\ President Donald Trump, including editorial and opinion content. Participants in r/science discuss recent press releases and papers announcing scientific advances. There, moderators enforce rules that emphasize the importance of referring to peer-reviewed scientific literature. Participants in r/news share links to breaking news, and engage in rapid-fire discussion as stories evolve over the course of minutes and hours, and moderators enforce a no-editorial rule.

To parallel our analysis of the debates and philosophical texts, we order the comments in each submission by time, \emph{i.e.}, we monitor the entire conversation as it evolves, even when the branching structure might serve to partially segregate the members involved.

Each source---book, debate, and submission---is thus brought into the same form: a stream of words. We then discretize this stream into $k$-word ``chunks'', with $k$ varying from 25 to 250. The semantic content of each chunk then serves as a marker of the position of the discussion at the corresponding point in time. Before chunking, we remove metadata, downcase all words and remove punctuation, and drop the fifteen most common words. In the case of the French-language texts, we drop accents, which helps mitigate errors due to OCR.

We use topic modeling to quantify the semantic content of the chunks~\cite{blei}. We treat each text as a collection of documents (chunks); the model then maps each chunk into a probability distribution over ``topics'' which summarize the content of the chunk in relation to the others. Because the topic model maps chunks into a \textit{probability} space, we can then use mathematical tools associated with Bayesian reasoning and information theory to quantify the relationship between one chunk and another. 
Fig.~\ref{word_dist} (appendix) shows the distribution of sentence lengths and post or speech lengths, after processing. Above $k$ equal to 100, the overwhelming majority of chunks contain at least three sentences, speeches, or posts; below that point, chunks begin to include single sentences or clauses. This range of $k$ thus allows us to track both how sentences are strung together and how paragraphs are knit into cohesive wholes.

A topic model maps the chunks of each source text to a common $N$-dimensional probability simplex; each chunk occupies a unique position in that space. We quantify the distance from one chunk to the next using the Kullback-Leibler (KL) divergence, or surprise,
\begin{equation}
\mathrm{KL}(\vec{p}, \vec{q}) = \sum_{i=1}^N p_i \log_2{\frac{p_i}{q_i}},
\end{equation}
where $\vec{p}$ is the probability distribution over topics for the current chunk, and $\vec{q}$ is the distribution for the chunk just previous. The value of $\mathrm{KL}(\vec{p}, \vec{q})$ corresponds to the Bayesian surprise~\cite{itti2009bayesian} upon encountering the current chunk, given a model of semantics built on the chunk just previous. \citeA{barron} and the Supplementary Information of \citeA{murdock2017exploration} provide a more extended account of the different ways in which Kullback-Leibler divergence can be interpreted as a property of individual-level cognition. 

Our choice of how to model the underlying semantic space of the discussion leads to a particular view on the texts as a whole. Two implications are worthy of particular attention. First, we have built our topic model for each source using only the chunks within that source. Informally, we allow the source to define the semantic space on its own terms. It is equally interesting to understand a text as embedded in a larger space; this was done by \citeA{joint_model}, where works by Charles Darwin and Alfred Russel Wallace were modelled in the context of the works that Darwin was known to have read. 

For example, our French Revolution source is modelled in the context of the speeches of that day, but it is just as possible to consider it in the context of all the speeches that month, or all the speeches before and to come. A philosophical text can be understood in terms of the topics it explores, or in terms of the overall project of the philosopher in question, or, indeed, the entire history of human thought.

Second, our definition of chunks is based on counting words. This has the advantage of simplicity. It is natural, if only approximate, to assume that a reader reads, or a listener hears, words at a roughly constant rate. Our choice also enables us to connect more directly to questions of how language is perceived and processed in the brain. 

An alternative choice, natural for written texts, is to delimit chunks using paragraph boundaries. This is often a means for a writer to signal the transition from one unit of meaning to another. A downside to this is that it can make the analysis particularly sensitive to what is often a minor detail of presentation. The insertion of a paragraph mark may be for stylistic reasons that have little to do with the content; online, users may have different standards for when to break up their text into separate remarks.

Another alternative is the use of a chunk size scaled to the source length, so that every source has the same number of chunks, and longer sources have larger chunks; one would see the \emph{Critique of Pure Reason} as having the same number of units as a day of the French Revolution.

Each of these choices has something to recommend it. For simplicity, in this work, we take the uniform chunk as our basis of analysis. In addition to its conceptual justification, it also has the advantage of avoiding the confounds introduced by variable chunk sizes both within and across texts: short chunks, for example, may be modeled differently than long ones, particularly when the length gets small.

\subsection{L\'{e}vy Flight Model}

The distribution of chunk-to-chunk surprise for each text puts constraints on the underlying process that generates each chunk from the previous ones. Here, we use a simple model that says that the position of the next chunk in semantic space is given by a combination of two constraints. First, the underlying stationary distribution, \emph{i.e.}, the space that the system ends up covering, including the biases in the system that tilt it to spend more time one some topics than others. Second, a preference for the level of surprise from the chunk just preceding, \emph{i.e.}, the extent to which the system makes short versus long leaps across the space it covers.

Our model is a simple biased random walk, where the distribution of jump sizes can have a very high degree of variance. In the physical sciences, these are known as L\'{e}vy Flights (see, \emph{e.g.}, \citeA{kleinberg2000navigation}), and are common to naturally occurring processes such as human movement and animal foraging~\cite{shlesinger2006random,reynolds2009levy}. Generic arguments show that L\'{e}vy Flights are optimal search processes in the case where the searching agent has limited ability to look beyond, and directly evaluate, a nearby radius~\cite{viswanathan1999optimizing}.\footnote{The formal definition of a L\'{e}vy flight often includes a power-law distribution of jump sizes, although this is not necessary; in our work, we require only a high variance and ``heavy-tailed'' distribution of jump sizes; the analogous processes in real space include not only the pure power-law case, but also a composite Brownian walk, with switching between large and small steps~\cite{benhamou2007many}.}

We adapt the L\'{e}vy Flight paradigm to the constraints of the probability simplex, where long jumps are determined not by an $L^n$ metric but the Kullback-Leibler divergence, a measure of cognitive surprise. In particular, given a particular location of chunk $i$, $\vec{v}_i$, we determine the position of the next chunk, $i+1$, by sampling from a Dirichlet distribution whose parameters are determined in part by the current location, and in part by the stationary distribution.

That stationary distribution is modeled by a Dirichlet distribution (formally, the posterior distribution given the Dirichlet prior and the documents themselves), parameterized by an $N$-dimensional vector, $\vec{\alpha}$. The position of a chunk $\vec{v}_{i+1}$ is then given by a draw from a Dirichlet that deviates from the stationary case, tilted to prefer locations nearby the chunk, $\vec{v}_i$, just previous,
\begin{equation}
\vec{v}_{i+1} \sim \mathrm{Dir}(\vec{\alpha}+\lambda \vec{v}_i).
\label{dir}
\end{equation}
Here, $\lambda$ plays the role of a focusing parameter. When $\lambda$ is large, a sample from the Dirichlet distribution in Eq.~\ref{dir} is very similar to $\vec{v}_i$; the next chunk is close to the one that came before. When $\lambda$ is very small, conversely, a sample tends to look like something drawn from the stationary distribution of the process as a whole (as a reminder, a draw from $\mathrm{Dir}(\vec{\alpha})$ corresponds to the posterior distribution of topics found by the topic model---a random draw of a ``typical'' chunk.) If the exploration of the space is completely uncorrelated (as might happen if the order of the chunks were shuffled, for example, in a null model), $\lambda$ would hover around zero.

\begin{figure}
  \centering
  \includegraphics[width=\linewidth]{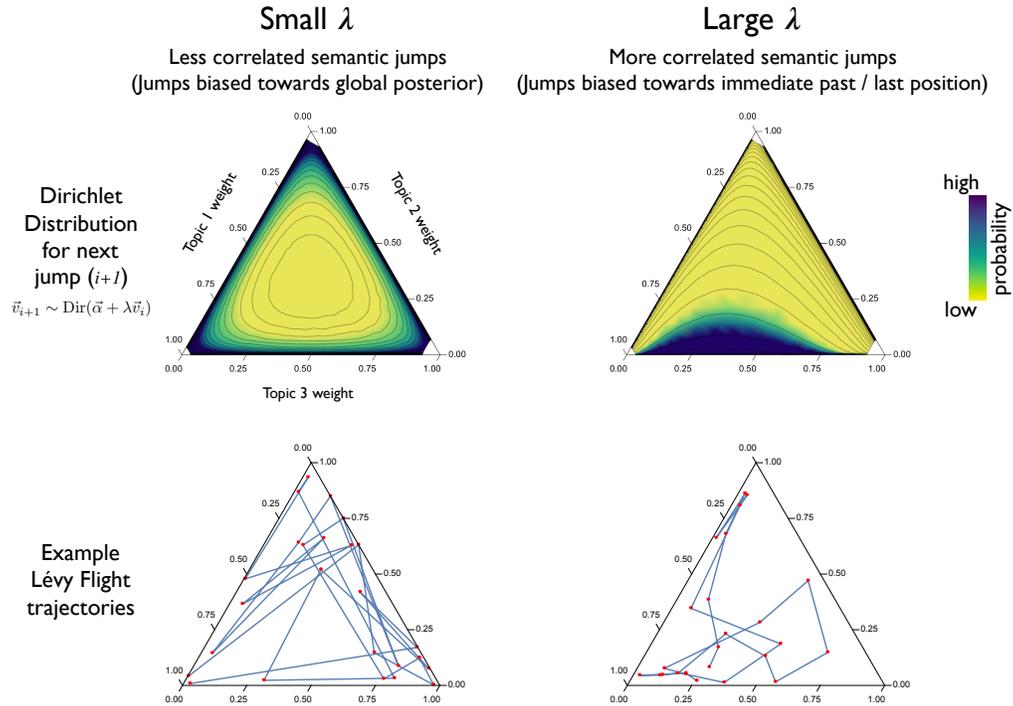}
  \caption{Long and short-range jumps on the simplex, here for the easily-visualized case of three topics. Top row: When $\lambda$ is close to zero (left panel), the next step is sampled from a distribution close to the stationary posterior, concentrated uniformly around the edges of the simplex (dark blue, higher probability) as is familiar for a standard topic model. When $\lambda$ is large (here, equal to $10$, a reasonably large excursion for philosophical texts on coarse-grained scales), the next step is sampled from a distribution strongly biased by the previous position (here, set equal to $\{0.38, 10^{-5}, 0.62\}$). Bottom row: left, twenty-five points in a sample trajectory with $\lambda$ set to zero, \emph{i.e.}, uncorrelated steps dictated by the posterior stationary distribution. Right, a second trajectory where $\lambda$ is drawn from a log-normal distribution with $\mu$ and $\sigma$ both set to unity, values similar to philosophy- and debate-like exploration patterns. Each subsequent step is, on average, closer to the previous one than in the uncorrelated case, with occasional long-range jumps intermingled with short steps.}
  \label{simplex}
\end{figure}
The effect of different values of $\lambda$ can been seen visually in Fig.~\ref{simplex}, where we show a toy model with three topics. When $\lambda$ is small, the system samples from the underlying stationary distribution; as $\lambda$ gets larger, the jumps are concentrated more and more around the original location.

To model the balance between focus and exploration, we take $\lambda$ itself, at each step, to be drawn from a (naturally heavy-tailed) log-normal distribution,
\begin{equation}
\lambda \sim \mathrm{LogN}(\mu, \sigma).
\label{logNdef}
\end{equation}
The parameter $\mu$ quantifies the average jump size, \emph{i.e.} the pace of the argument. The parameter $\sigma$ quantifies the variance in that jump size, \emph{i.e.}, the burstiness of the argument. We determine $\mu$ and $\sigma$, along with error bars for each, empirically from the data. When $\mu$ is large, texts are more focused; when $\mu$ is small, texts tend to jump more randomly from topic to topic. Large $\sigma$ complicate these effects; if $\mu$ is small, but $\sigma$ is large, then a text is usually unfocused, but occasionally very highly focused when a chance event leads to an anomalously large $\lambda$. Other heavy-tailed distributions are possible.\footnote{We use the log-normal as our heavy-tailed distribution for two reasons. First, in contrast to other functions such as the Pareto distribution, the log-normal allows us to vary the median and variance independently of each other. Second, it has a natural interpretation in terms of a central limit theorem for multiplicative gains. This makes it possible to connect to simple process models for the underlying phenomenology---although we do not propose one here.}

For each text, we fit the $\mu$ and $\sigma$ parameters to the distribution of chunk-to-chunk surprises. We do this in a Bayesian fashion (see Appendix~\ref{app_bayes}) that allows us to determine error bars on our estimates. Optimized C code {\tt levytopic} to implement these fits on arbitrary distributions, and is available at \url{http://santafe.edu/\~{}simon}.

\section{Results}

Our three collections, corresponding to the different levels of social organization, are naturally interpreted in the context of the others. We consider the one-to-many case (philosophical texts), the linear case (parliamentary debates), and the branching case (Reddit discussion), in turn. In each case, we look at how the values of $\mu$ and $\sigma$ change, or ``flow'', as we increase the coarse-graining scale $k$.

\subsection{Philosophical Texts}

\begin{figure}
  \centering
  \begin{tabular}{cc}
  \includegraphics[width=0.45\linewidth]{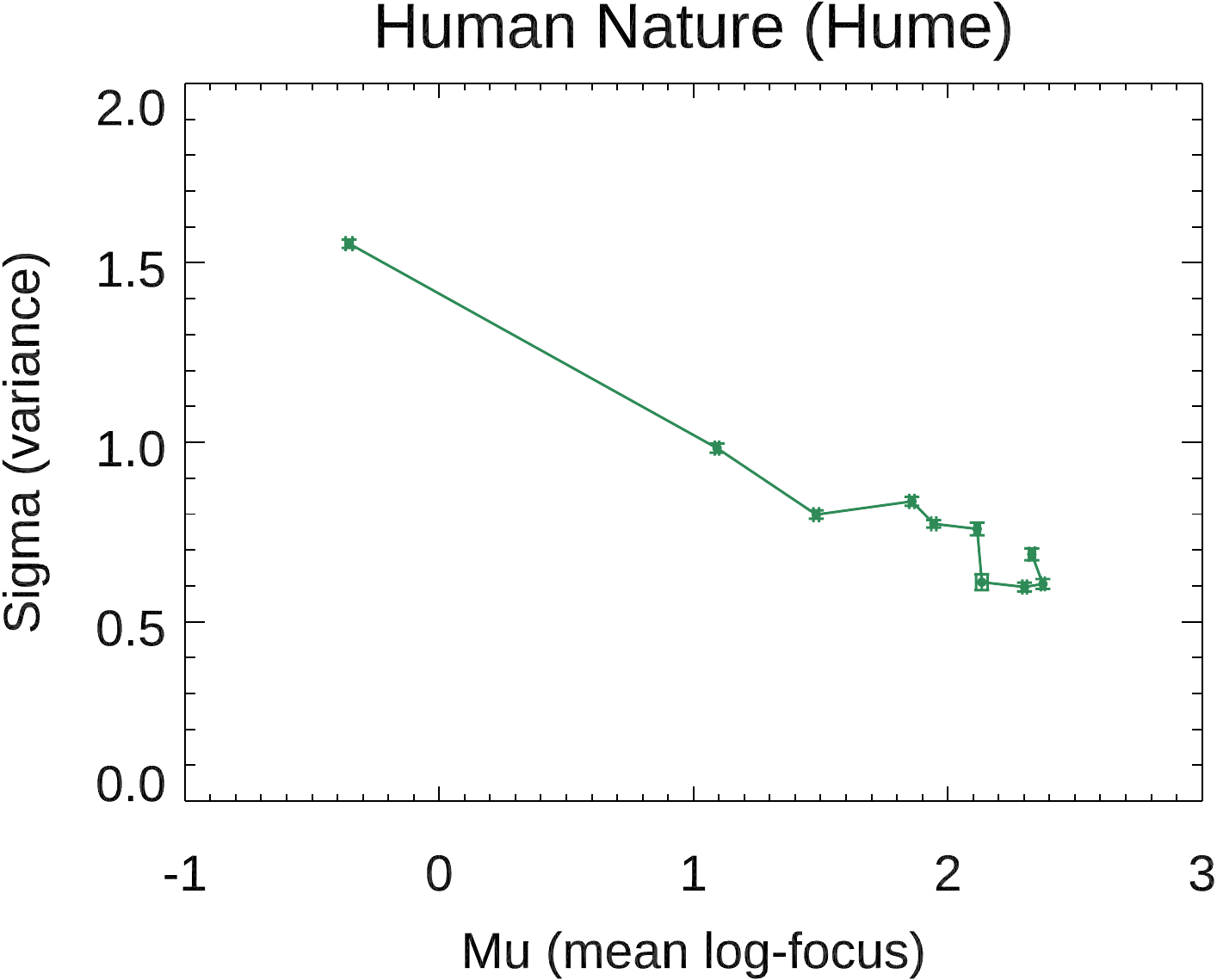} & \includegraphics[width=0.45\linewidth]{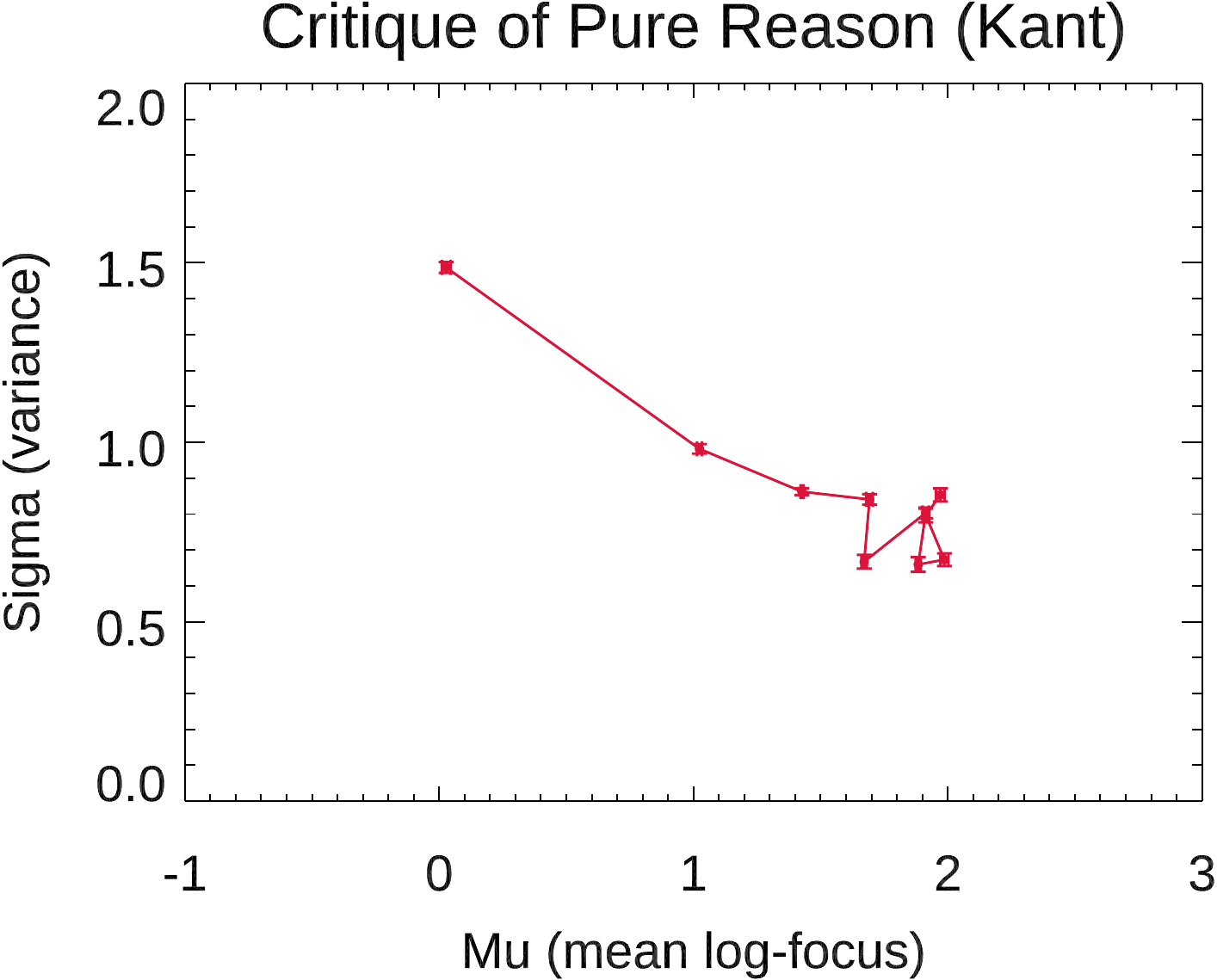}
  \end{tabular}
  \caption{Derived values of $\mu$ (the average of the logarithm of the focus parameter $\lambda$, Eq.~\ref{logNdef}) and $\sigma$ (the variance) as a function of scale, $k$, for David Hume's \emph{Treatise on Human Nature} and Immanuel Kant's \emph{Critique of Pure Reason}. The coarse-graining scale $k$ ranges from 25 to 250 words (after stop-word removal). The left-most point in each panel corresponds to the smallest scale. As $k$ increases, and we take texts in larger and larger units, the estimated $\mu$ parameter becomes larger, indicating higher levels of focus and prior-step domination.}
  \label{phil_example}
\end{figure}

Fig.~\ref{phil_example} shows the derived values of $\mu$ and $\sigma$ for the case of Hume's \emph{Treatise on Human Nature} and Kant's \emph{Critique of Pure Reason}. In both cases, the effect of coarse-graining is to reveal structure: as we more to larger and larger scales (from $k$ equal to 25 words, the left-most point on each plot, to $k$ equal to 250 words, on the right), the focus parameter $\lambda$ turns out to be drawn from a distribution biased towards larger values. Table~\ref{phil_table} shows this in a different way, showing the median, and ranges, of $\lambda$ given the $\mu$ and $\sigma$ parameters. In the case of the \emph{Treatise of Human Nature}, for example, the median value of $\lambda$ moves from $0.81$ at the smallest scales ($k$ equal to 25) to $8.11$ on the largest scales ($k$ equal to 250), with excursions as large as $19$, a level of focus similar to the shift from the left to the right panel of the demonstration figure, Fig.~\ref{simplex}.

\begin{table}[h!]
  \centering
  \begingroup
\setlength{\tabcolsep}{10pt} 
\renewcommand{\arraystretch}{1.45} 
\begin{tabular}{l|l|l}
    Text & $\lambda$ range, $k=25$ & $\lambda$ range, $k=250$ \\ 
    & (Small scale) & (Large scale) \\ \hline
    Aristotle, Metaphysics & $1.48_{~0.38}^{~5.73}$ & $10.56_{~4.62}^{~24.13}$ \\
    '', Ethics & $1.38_{~0.27}^{~6.96}$ & $5.82_{~2.83}^{~11.99}$ \\  '', Rhetoric & $1.14_{~0.18}^{~7.38}$ & $8.08_{~3.74}^{~17.46}$ \\
    Hume, Understanding & $0.70_{~0.14}^{~3.51}$ & $10.28_{~5.04}^{~20.97}$ \\
    '', Human Nature & $0.81_{~0.15}^{~4.23}$ & $8.11_{~3.46}^{~19.01}$ \\
    '', Morals & $0.45_{~0.06}^{~3.15}$ & $4.23_{~1.74}^{~10.32}$ \\
    Kant, Pure & $1.03_{~0.22}^{~4.80}$ & $7.17_{~2.96}^{~17.37}$ \\
    '', Practical & $0.92_{~0.18}^{~4.84}$ & $6.58_{~2.58}^{~16.78}$ \\
    '', Judgement & $1.30_{~0.29}^{~5.82}$ & $10.72_{~4.67}^{~24.56}$ \\
  \end{tabular}
\endgroup
  \caption{The range of focus parameters $\lambda$, for the nine philosophical texts. Shown are the median, and, in sub- and super-script, respectively, the $15\%$ and $85\%$ points of the cumulative distribution. At the finest-grained scales, textual dynamics are usually only mildly influenced by the current position within topic space, although occasional high-focus sequences may appear. At the most coarse-grained scales the system is much more strongly determined by the previous location, indicating the emergence of large-scale structure.}
  \label{phil_table}
\end{table}

\begin{figure}
  \centering
  \includegraphics[width=0.65\linewidth]{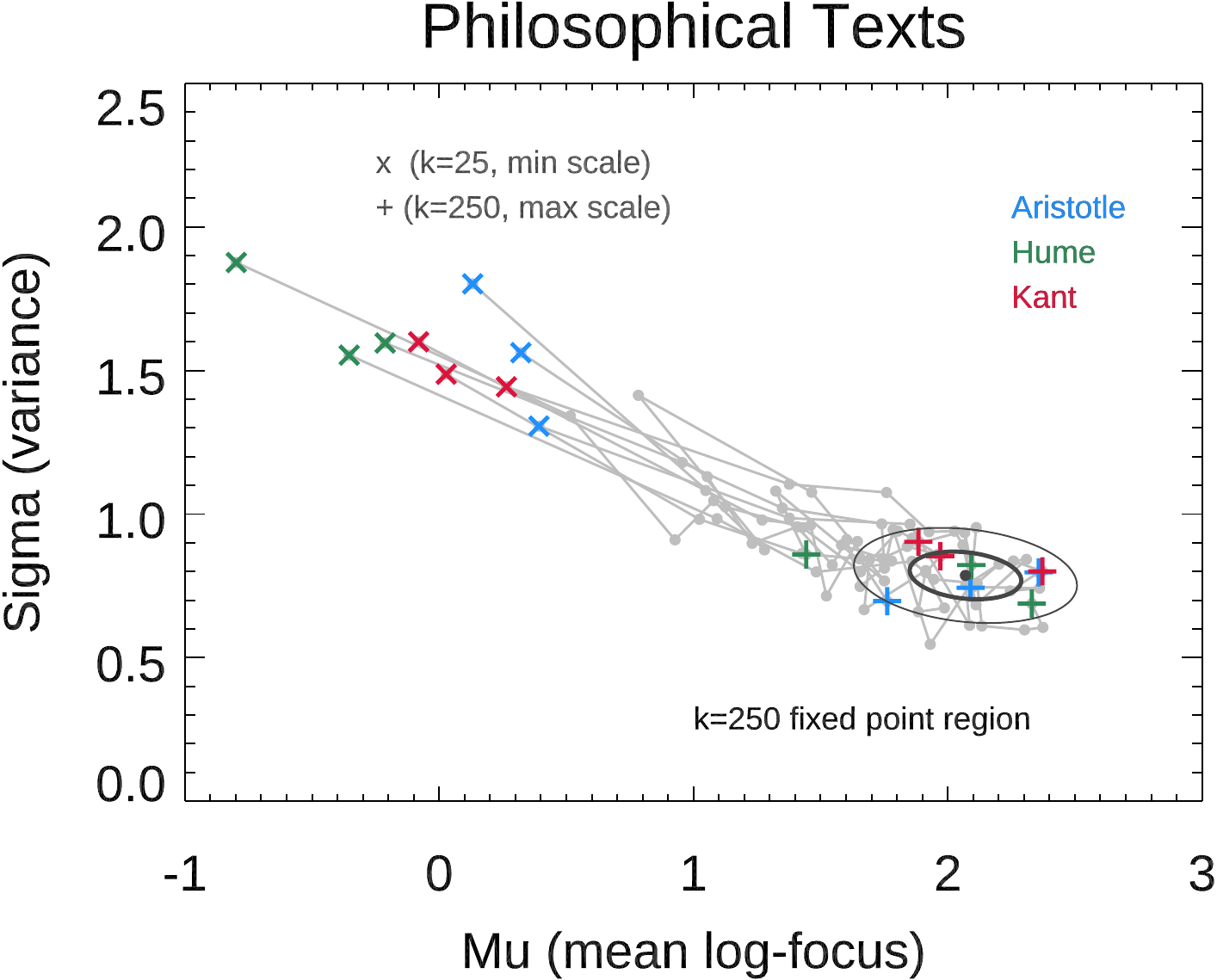}
  \caption{All nine philosophical texts, from Aristotle, Hume, and Kant, coarse-grain to a limit region with a centriod of $\{\mu=2.07, \sigma=0.79\}$ . With one exception (Hume's \emph{Morals}), all points at $k=250$ lie within the $2\sigma$ contours.}
  \label{phil_joint}
\end{figure}

The limit regions for both Aristotle and Kant lie in a similar portion of the space, around $\mu$ of $2.0$ and $\sigma$ of $0.75$. It so turns out that all nine of the philosophical texts under consideration coarse-grain to a tightly-bounded region in this range, as can be seen in Fig.~\ref{phil_joint}. (One text, Hume's $\emph{Ethics}$ enters this region for a period around $k$ equal to 200, but at the very largest scales ends up outside the computed $2\sigma$ contours of this region, potentially indicating some additional structure.)

\subsection{Debates of the French Revolutionary Parliament}

\begin{figure}
  \centering
  \begin{tabular}{cc}
\includegraphics[width=0.45\linewidth]{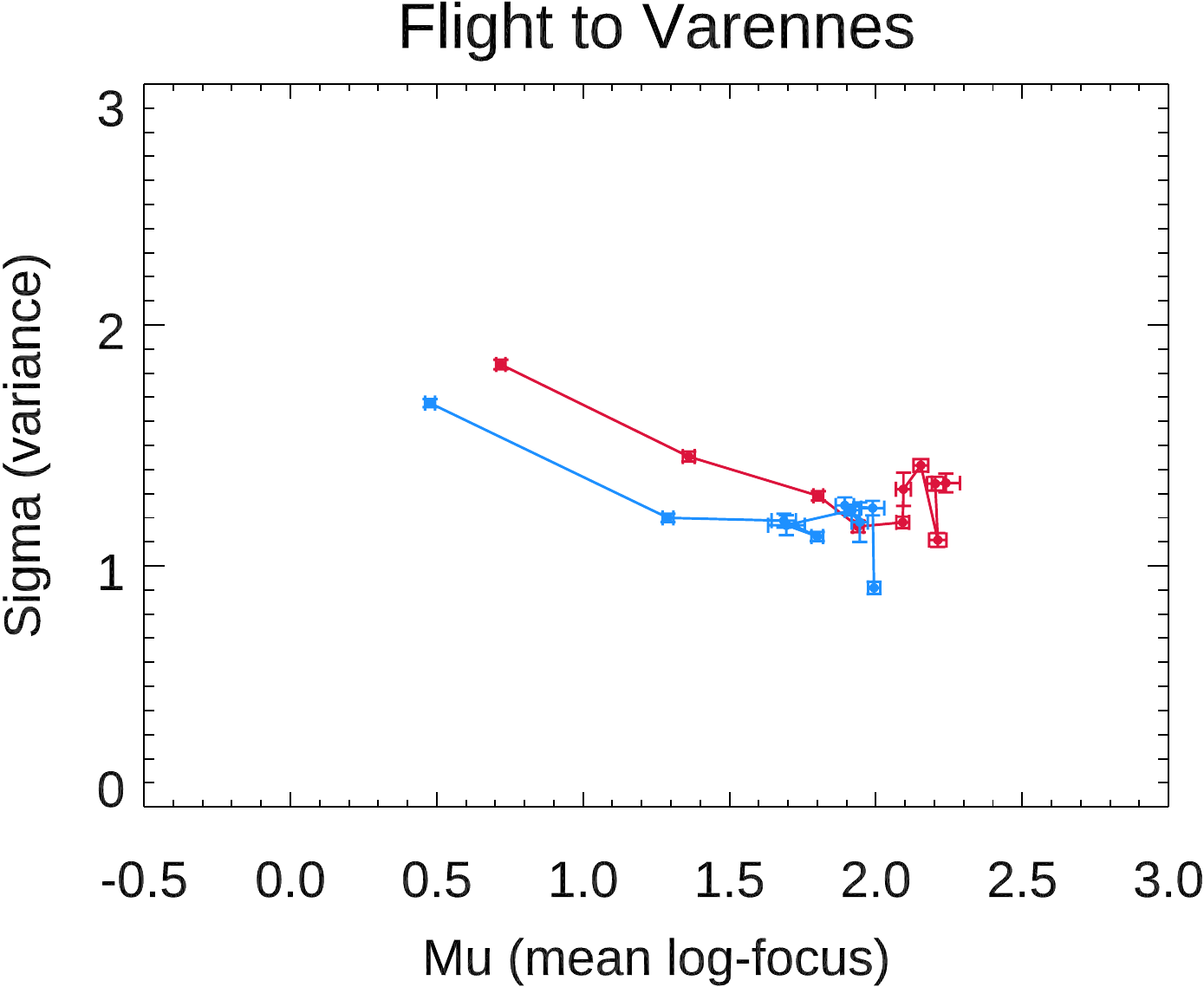} & \includegraphics[width=0.45\linewidth]{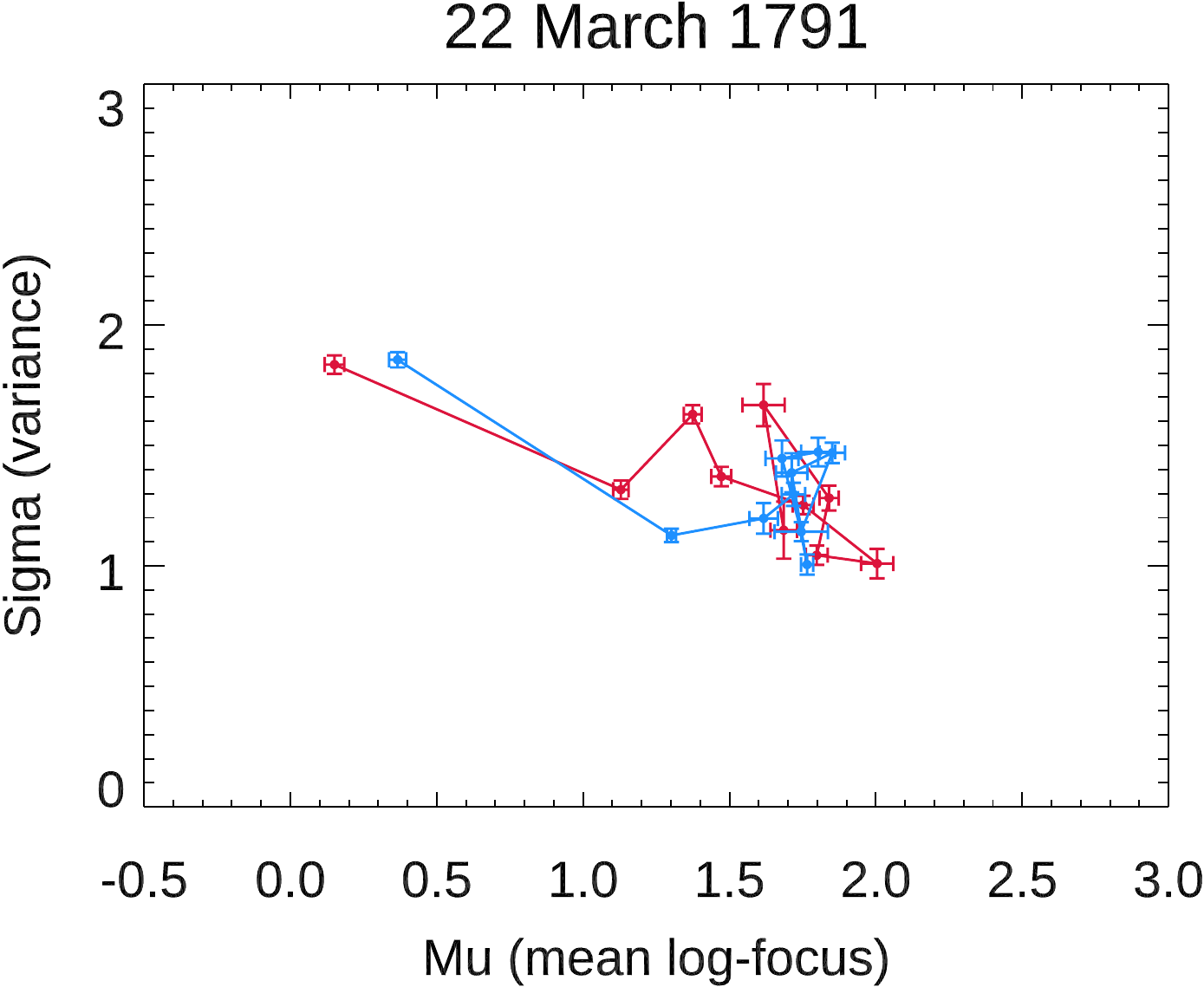} \\
  \vspace{0.5cm} \includegraphics[width=0.45\linewidth]{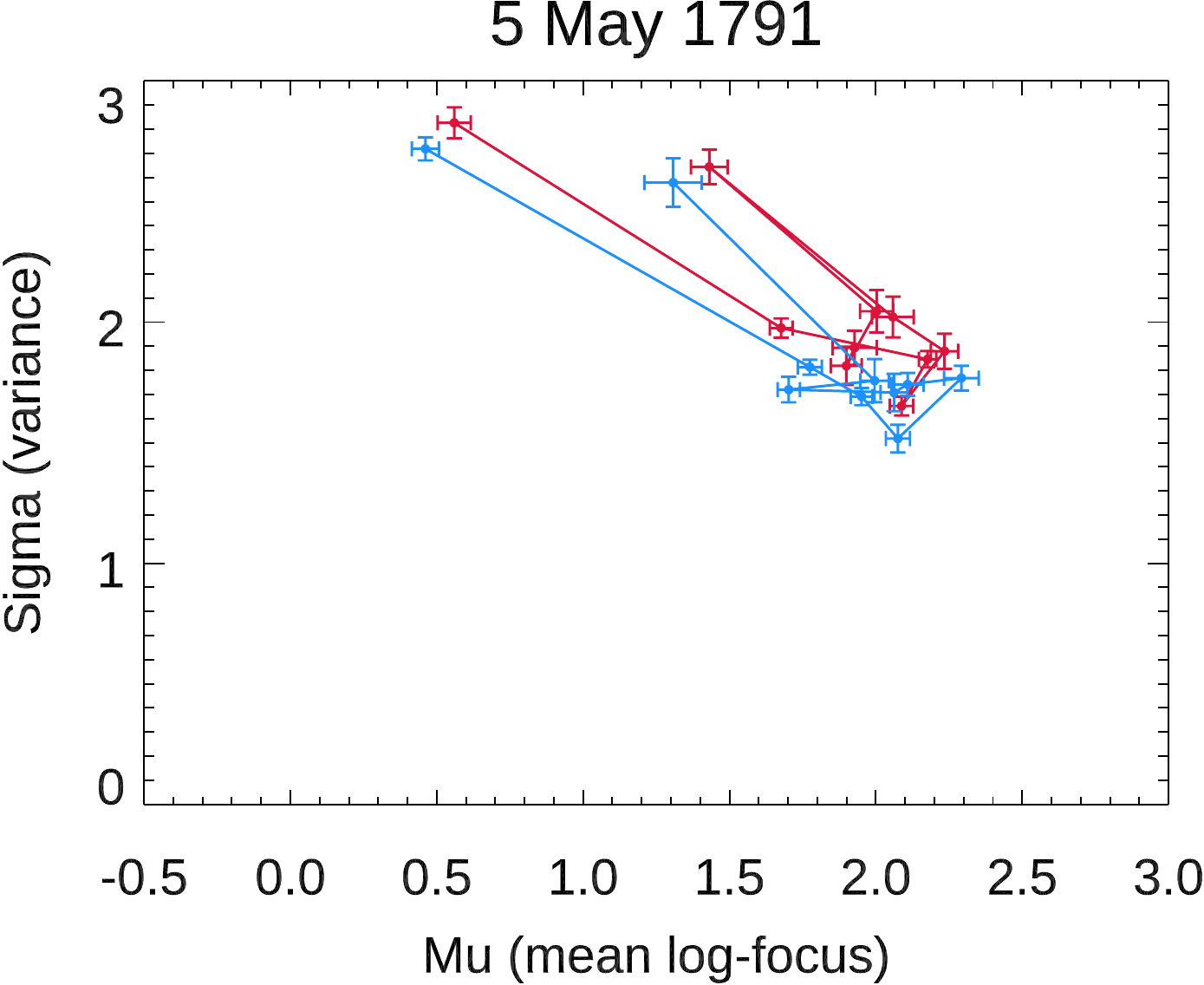} & \includegraphics[width=0.45\linewidth]{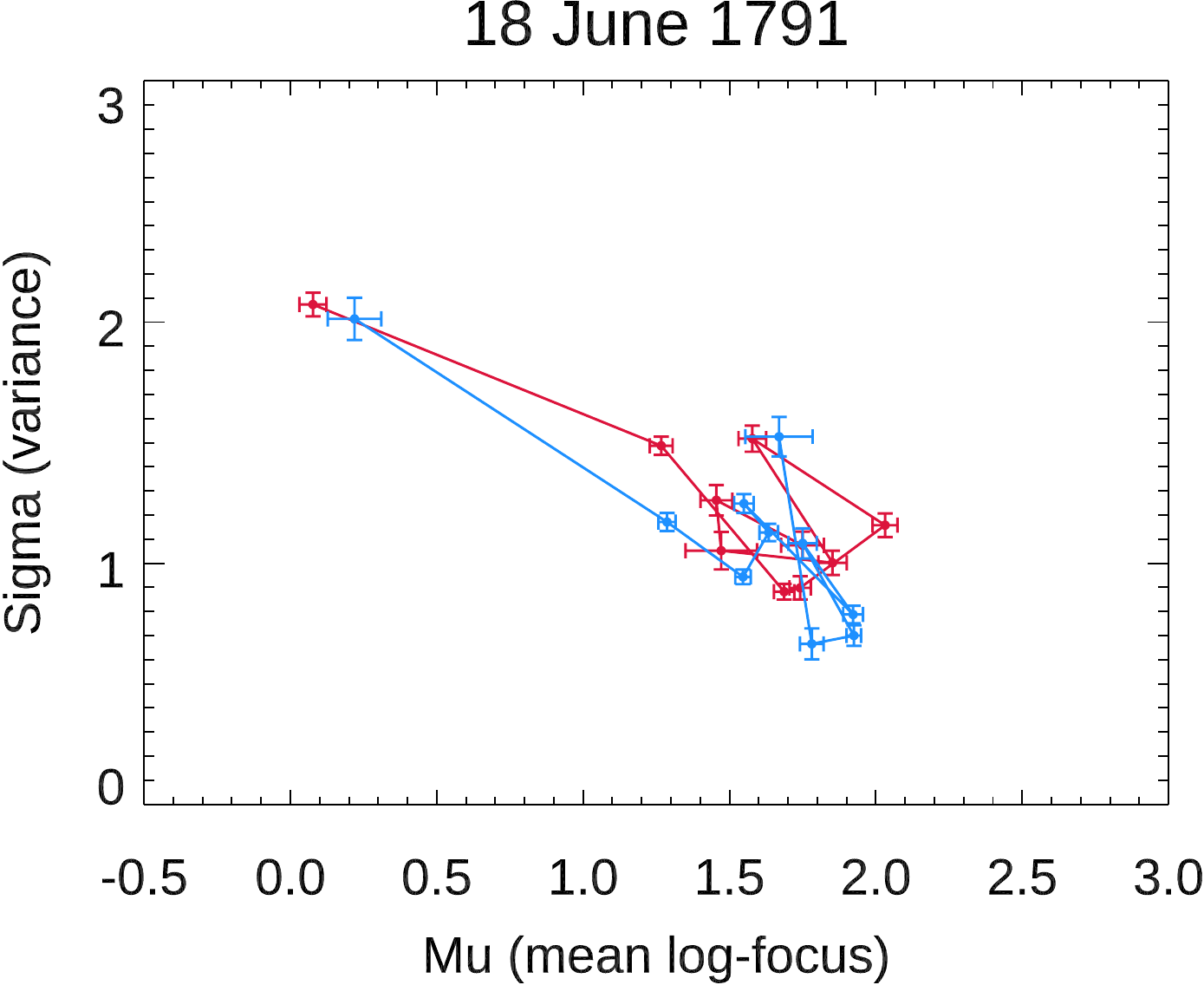}  \end{tabular}
  \caption{The $\mu$-$\sigma$ flow for four days of debate in the French Revolutionary Parliament. The parameter values for the original French text are shown in red, and the machine-produced translation to English in blue. The way the parameters move to lower $\sigma$, and higher $\lambda$, is robust to the shift in language. One day, 5 May 1791, shows a significantly higher sigma value across all scales; this is discussed in the main text.}
  \label{frev}
\end{figure}

Fig.~\ref{frev} shows the derived values of $\mu$ and $\sigma$ for the four days of the French Revolution. The panels show the agreement between an analysis conducted on the original French text, and a machine-produced translation into English. The semantic structure detected by the topic model, and the dynamical flows over that structure, are independent of language.

\begin{figure}
  \centering
  \includegraphics[width=0.65\linewidth]{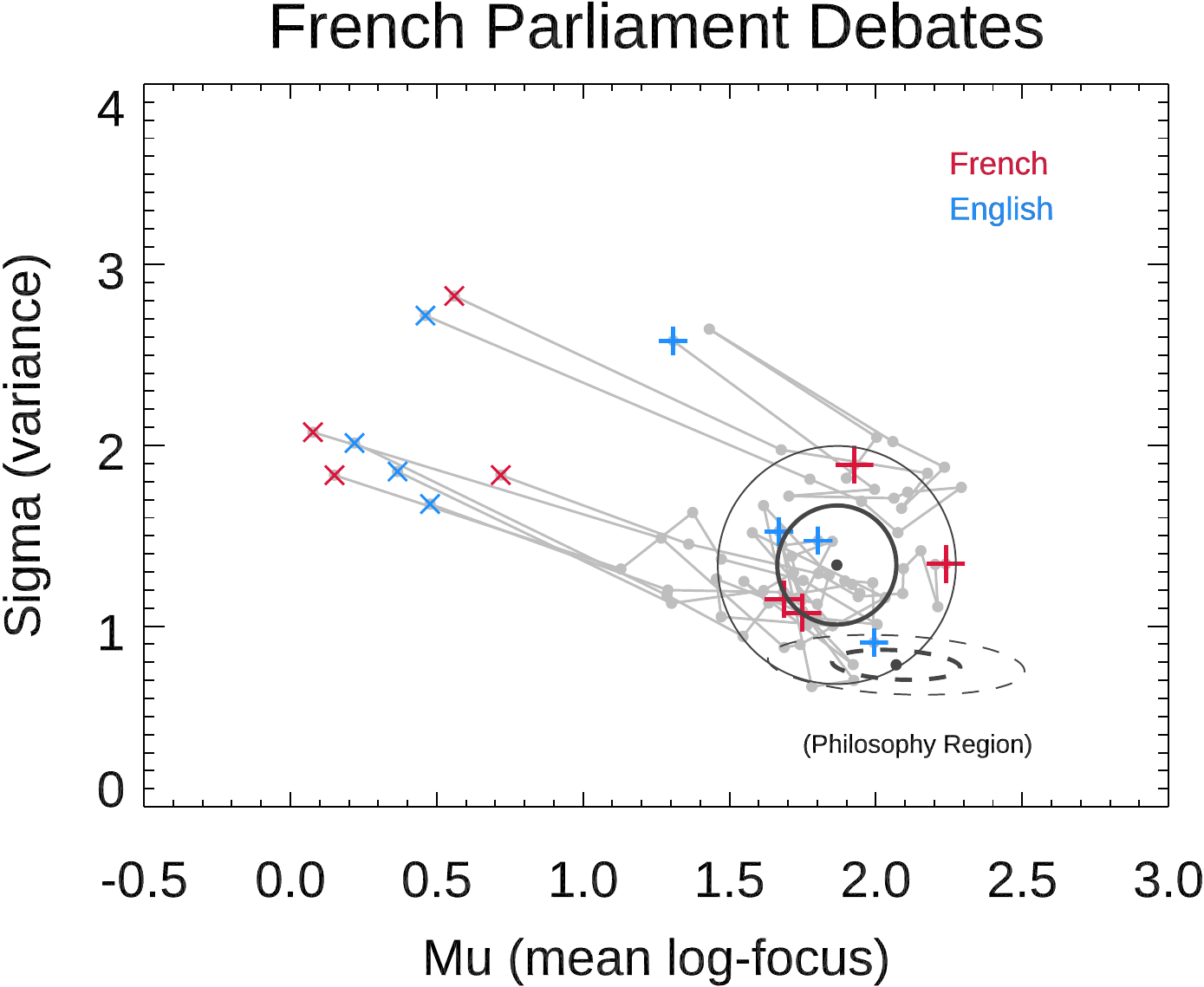}
  \caption{Parliamentary debates, whether in French or English, coarse-grain to a limit region with a centroid of $\{\mu=1.87, \sigma=1.34\}$. This is higher in variance than the region associated with philosophical texts, and corresponds to greater swings between short- and long-range jumps. The $k=250$ region for both parliamentary debates, in both French and English, is shown by the solid contours, with the philosophy region overlaid with dashed contours for comparison. The overall pattern, of increasingly focused discussion on larger scales, follows that seen in the philosophical texts.}
  \label{frev_joint}
\end{figure}

\begin{table}[h!]
  \centering
  \begingroup
\setlength{\tabcolsep}{10pt} 
\renewcommand{\arraystretch}{1.5} 
  \begin{tabular}{l|l|l}
    Text & $\lambda$ range, $k=25$ & $\lambda$ range, $k=250$ \\ 
    & (Small scale) & (Large scale) \\ \hline
    Flight to Varennes & $2.05_{~0.31}^{~13.76}$ & $9.39_{~2.33}^{~37.80}$ \\
    '' (English) & $1.61_{~0.28}^{~9.15}$ & $7.34_{~2.86}^{~18.84}$ \\
    22 March 1791 & $1.16_{~0.17}^{~7.79}$ & $5.39_{~1.64}^{~17.74}$ \\
    '' (English) & $1.44_{~0.21}^{~9.87}$ & $6.07_{~1.32}^{~27.93}$ \\
    5 May 1791 & $1.75_{~0.09}^{~32.77}$ & $6.87_{~0.97}^{~48.92}$ \\
    '' (English) & $1.59_{~0.09}^{~26.55}$ & $3.70_{~0.26}^{~53.54}$ \\
    18 June 1791 & $1.08_{~0.13}^{~9.26}$ & $5.75_{~1.89}^{~17.52}$ \\
    '' (English) & $1.24_{~0.15}^{~10.03}$ & $5.31_{~1.09}^{~25.78}$ \\
  \end{tabular}
  \endgroup
  \caption{The range of focus parameters $\lambda$, for four days of debate in the French Revolutionary Parliament. As in the case of the philosophical texts, the median focus increases as one goes to larger scales, indicating the emergence of large-scale patterns of argument. At both small and large scales, argument dynamics show greater variability, with occasional moments of both high focus and long-range jumps. The overall patterns are independent of whether the analysis is conducted in the original French, or in a rough translation produced by Google Translate. The anomalously high variance for the 5th of May is due to the presence of a segment of particularly homogeneous content that contrasts with the more ordinary ebb and flow of debate.}
  \label{frev_table}
\end{table}
Fig.~\ref{frev_joint} overlays all eight curves (four days, and two languages), and the limit region for $k=250$. While debates show a similar level of mean log-focus as the philosophical texts considered in the previous section, the debates show higher variance. Step by step, the parliament shows similar levels of focus, but has excursions in both directions: to arbitrary leaps governed largely by the stationary distribution, and to highly-focused moments where the next chunk is almost entirely dominated by the previous step. This can also be seen in a comparison of Table~\ref{frev_table} with Table~\ref{phil_table}.

One day stands out as an exception to the overall pattern: the 5th of May, 1791, which has an anomalously high $\sigma$ (variance in focus). Inspection shows that this day has about 20\% of its content devoted to a listing of the location and management of church estates. A focus on a single set of repeating word patterns leads to low jump sizes in that period; at the end of those lists, the chamber returns to a more ordinary flow of topics and speakers. As one expects, the shift between unusually uniform content (very high focus) and ordinary debate (lower focus) is reflected in the higher $\sigma$ value, and in the $\lambda$ range of Table~\ref{frev_table}.

\subsection{Reddit Discussion Trees}

\begin{figure}
  \centering
  \begin{tabular}{cc}
    \includegraphics[width=0.45\linewidth]{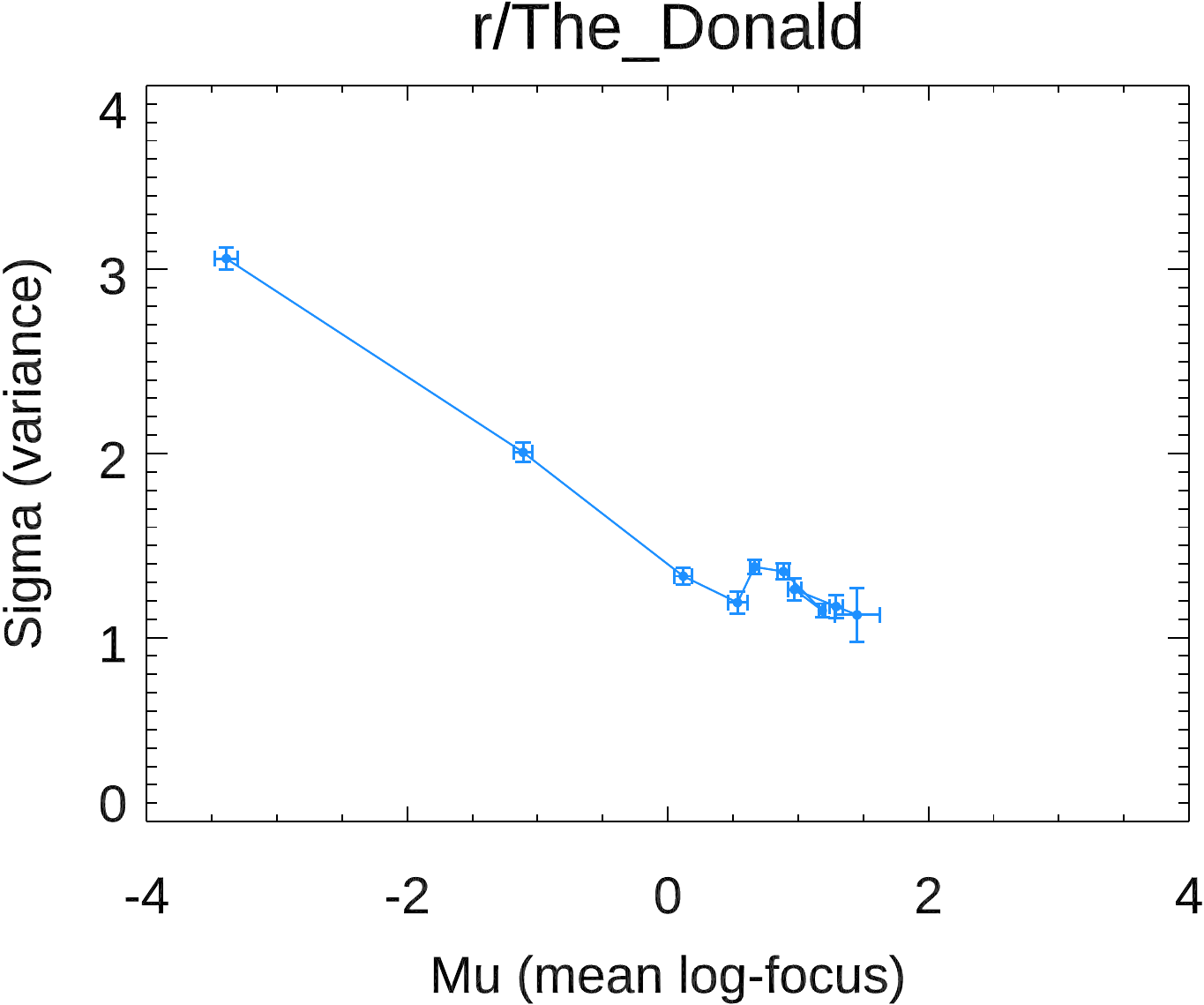} &     \includegraphics[width=0.45\linewidth]{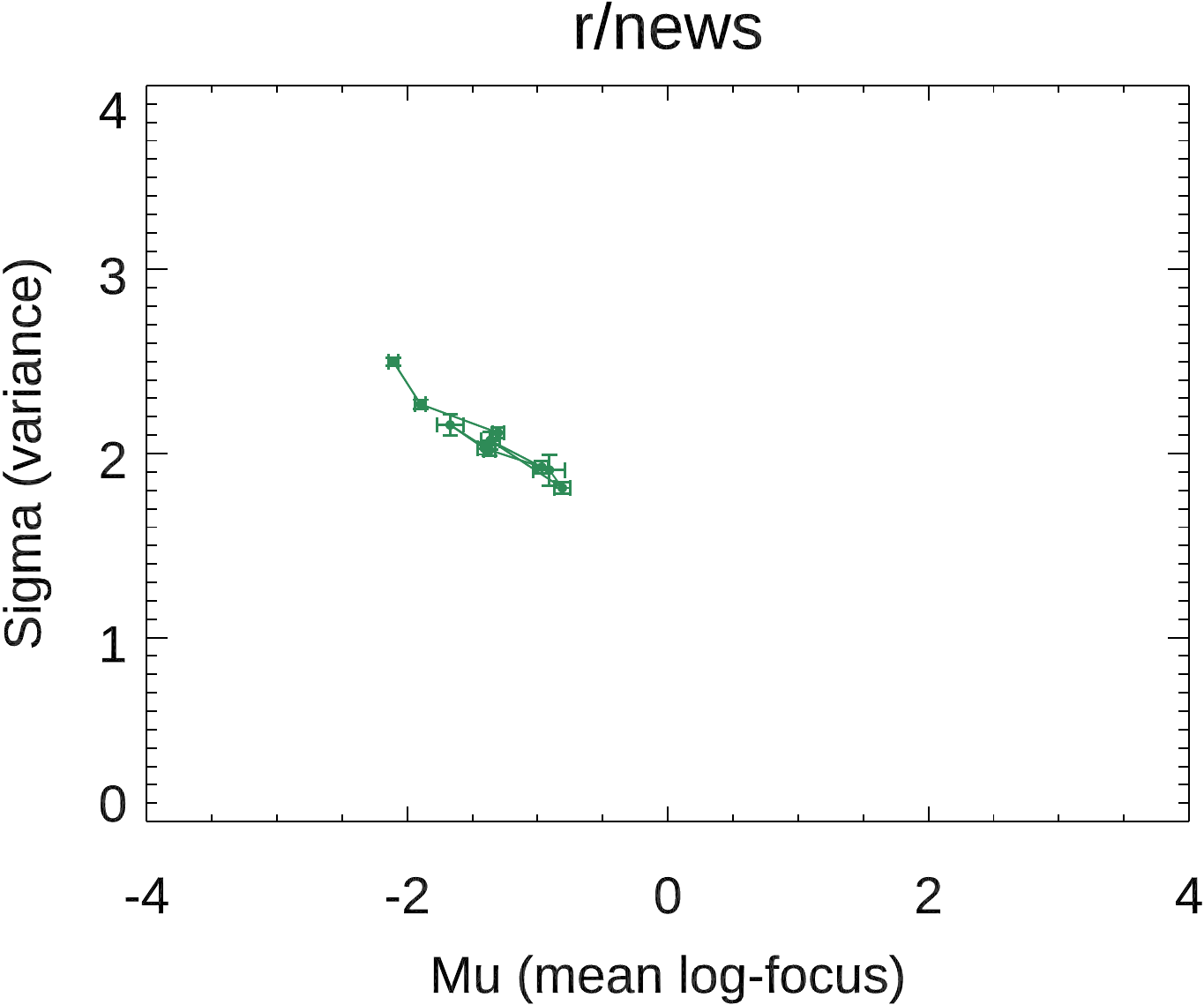} \\
        \includegraphics[width=0.45\linewidth]{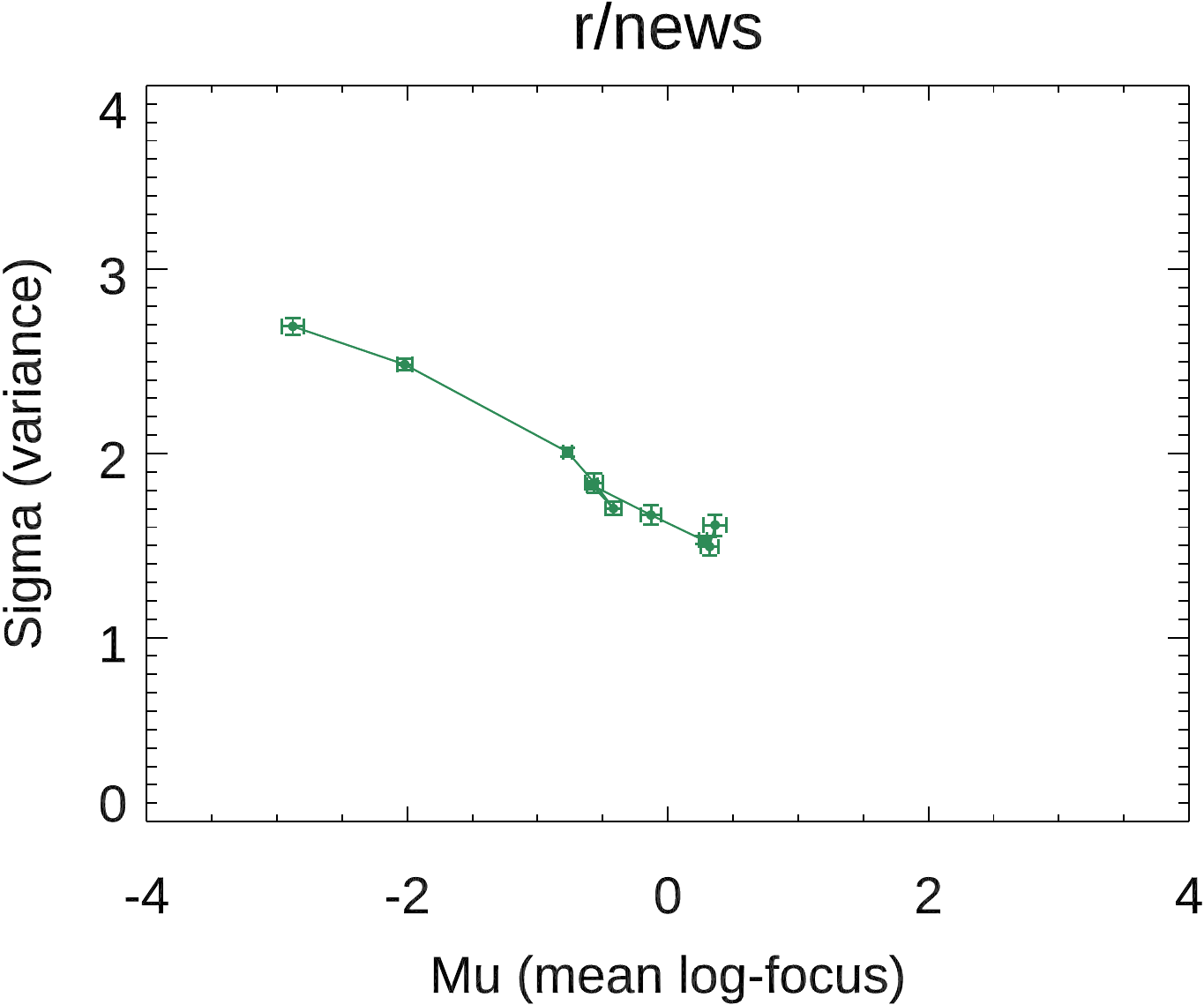} &     \includegraphics[width=0.45\linewidth]{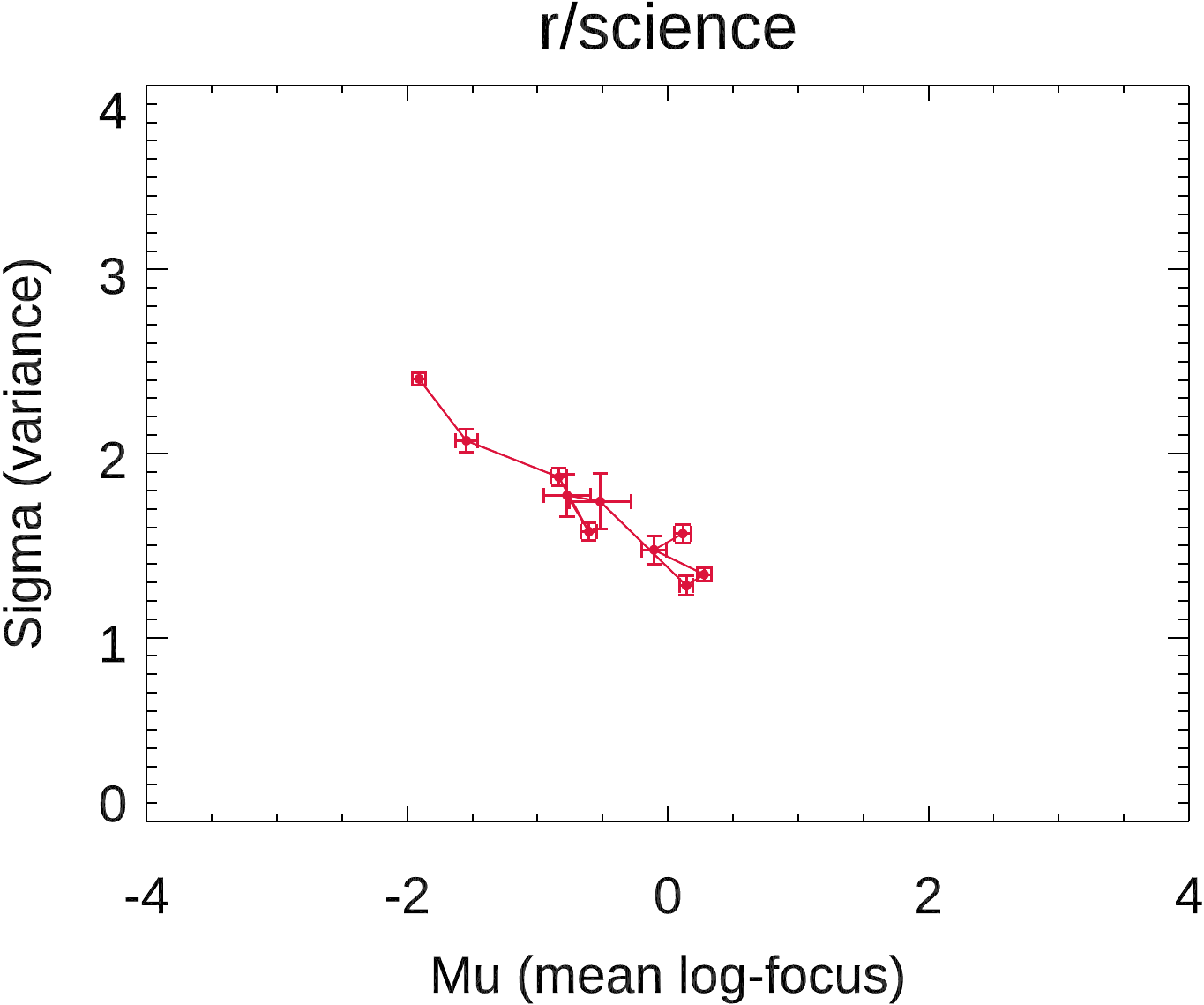}
          \end{tabular}
  \caption{Derived $\mu$ and $\sigma$ values for four submissions drawn from r/The\_Donald, r/science, and r/news. The overall pattern of flow from low focus, high sigma to high focus, low sigma follows the those established in the philosophical texts and parliamentary debates. Online discussions, however, tend to converge to lower values of $\mu$. They are also more likely to show reversals, \emph{i.e.}, the apparent loss of predictability as one goes to larger scales.}
  \label{reddit_example}
\end{figure}
\begin{figure}
  \centering
  \includegraphics[width=0.65\linewidth]{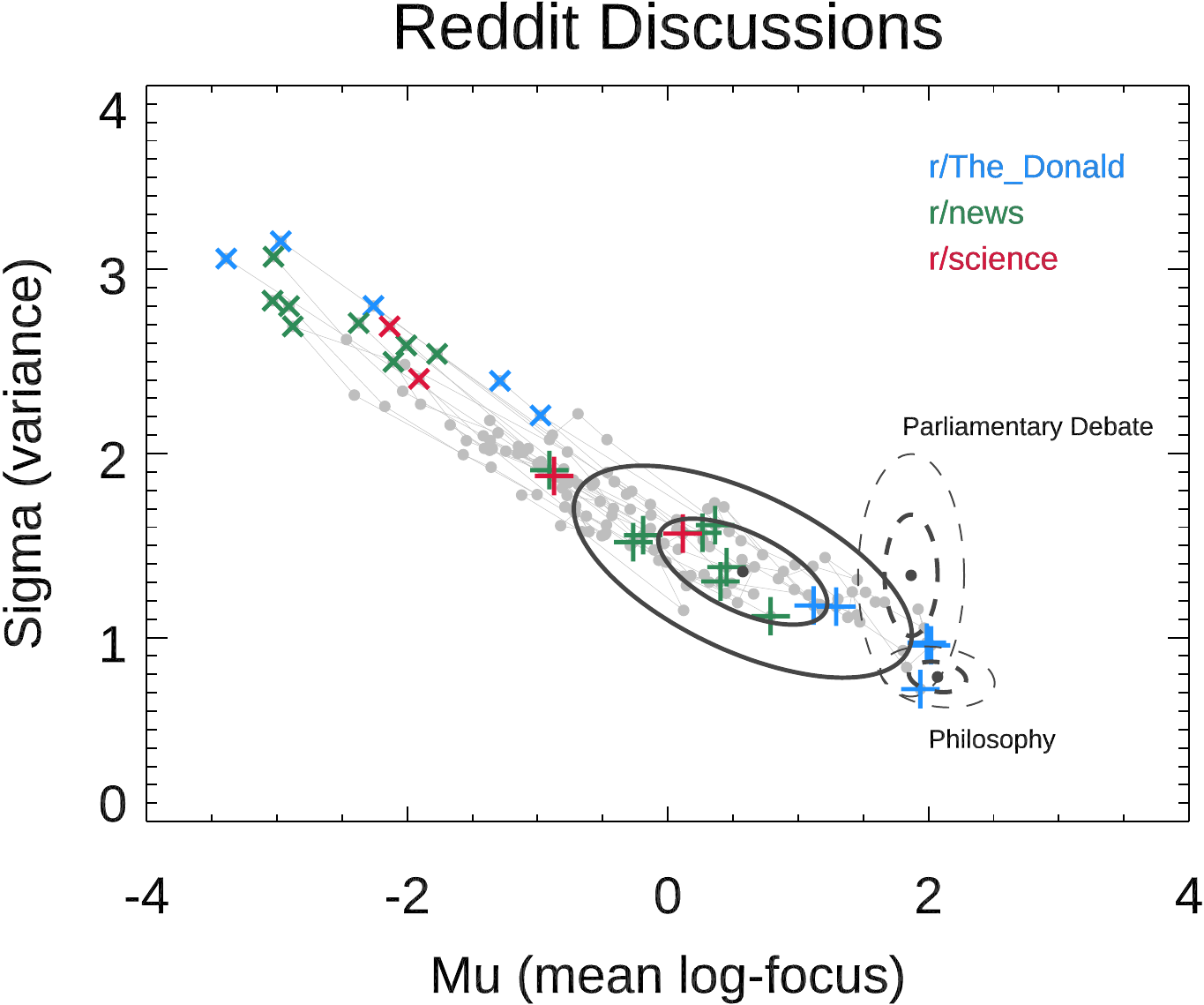} 
    \caption{All fifteen discussions from subreddits r/The\_Donald, r/science, and r/news that are analyzed in this paper. Overall, Reddit discussions coarse-grain to regions with significantly higher $\sigma$ values, and lower $\mu$ values than the philosophy and debates cases which are shown for comparison, as dashed black ellipses. Reddit discussions are less focused in the median, but swing more dramatically between long-range jumps and highly-focused steps. Political discussion on r/The\_Donald is the most debate-like (blue $+$ marks), while r/news and r/science (green and red $+$ marks) appear to exchange temporal ordering for more sophisticated structures enabled by Reddit's commenting software.}
  \label{reddit_joint}
\end{figure}

Fig.~\ref{reddit_example} shows examples of the $\mu$ and $\sigma$ parameters for the three subreddits r/science, r/news, and r/The\_Donald. The same flow from high-sigma, low-mu to low-sigma, high-mu appears. Both initial and final values of the parameters are shifted up and to the left compared to the text and debate cases. The coarse-graining process is in some cases more jagged: on going to larger scales some previous predictability can be lost. Fig.~\ref{reddit_joint} combines all fifteen submissions to show the overall pattern. 

\begin{figure}
  \centering
  \includegraphics[width=0.65\linewidth]{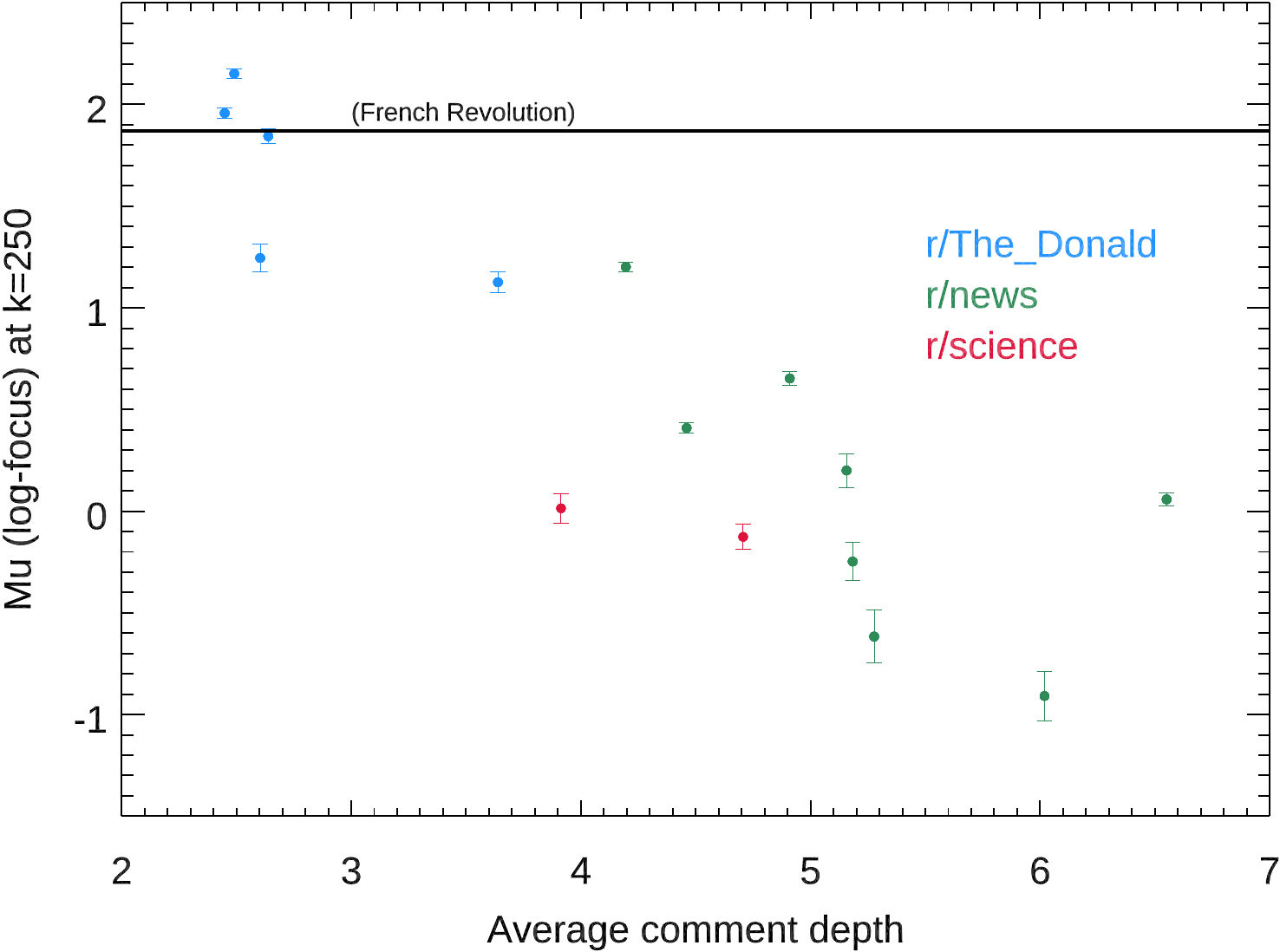} 
  \caption{The depth of comment trees partially predicts discussion focus, $\mu$, on large scales. When users create longer comment chains, they can fragment the structured linear development of the conversation, in contrast to what is found in parliamentary debates. Subreddit-level norms still matter, however: submissions on different subreddits can have very different levels of focus despite having similar depths.}
  \label{reddit_depth}
\end{figure}

On large scales, submissions from r/The\_Donald are the most similar to the debate and philosophy cases, suggesting that interactions on the forum develop linearly, in a way more determined by temporal order. Meanwhile, r/news and r/science show low $\mu$ and high $\sigma$ values well outside the text and debate ranges, even on the largest scales. This apparent lack of structure is most naturally explained by how participants use the branching potential of the commenting system to explore different aspects of the topic simultaneously. The hypothesis is partially borne out by the results shown in Fig.~\ref{reddit_depth}. There, we show how increasing the average depth of a comment in a submission partially predicts decreased $\mu$. The more deeply nested comments are, on average, the less focused the discussion appears to be from the point of view of a linear order.

\begin{table}[h!]
  \centering
  \begingroup
\setlength{\tabcolsep}{10pt} 
\renewcommand{\arraystretch}{1.5} 
  \begin{tabular}{l|l|l}
    Text & $\lambda$ range, $k=25$ & $\lambda$ range, $k=250$ \\ 
    & (Small scale) & (Large scale) \\ \hline
    r/The\_Donald & $0.28_{~0.02}^{~3.30}$ & $6.94_{~3.29}^{~14.65}$ \\
    ''  & $0.38_{~0.04}^{~3.71}$ & $7.30_{~2.66}^{~19.99}$ \\    
    '' & $0.03_{~<10^{-3}}^{~0.80}$ & $3.64_{~1.08}^{~12.22}$ \\
    '' & $0.10_{~<10^{-3}}^{~1.91}$ & $7.54_{~2.79}^{~20.34}$ \\
    '' & $0.05_{~<10^{-3}}^{~1.35}$ & $3.06_{~0.91}^{~10.34}$ \\
    r/news & $0.05_{~<10^{-3}}^{~1.0}$ & $0.83_{~0.16}^{~4.15}$ \\
    '' & $0.05_{~<10^{-3}}^{~0.90}$ & $2.20_{~0.69}^{~7.00}$ \\
    '' & $0.09_{~<10^{-3}}^{~1.55}$ & $0.77_{~0.16}^{~3.70}$ \\
    '' & $0.06_{~<10^{-3}}^{~0.92}$ & $1.44_{~0.27}^{~7.63}$ \\
    '' & $0.12_{~<10^{-3}}^{~1.62}$ & $0.40_{~0.06}^{~2.92}$ \\
    '' & $0.17_{~0.01}^{~2.37}$ & $1.30_{~0.26}^{~6.63}$ \\
    '' & $0.05_{~<10^{-3}}^{~1.17}$ & $1.57_{~0.37}^{~6.58}$ \\
    '' & $0.13_{~<10^{-3}}^{~1.96}$ & $1.50_{~0.39}^{~5.79}$ \\
    r/science & $0.12_{~<10^{-3}}^{~1.92}$ & $0.42_{~0.06}^{~2.93}$ \\
    '' & $0.15_{~0.01}^{~1.79}$ & $1.12_{~0.22}^{~5.68}$ \\
  \end{tabular}
  \endgroup

  \caption{The range of focus parameters $\lambda$, for fifteen Reddit submissions. At the smallest scales, the dynamics are often very close to completely unstructured leaps at random across the semantic space ($\lambda\ll 1$). On larger scales, we see (to a lesser extent) the same recovery of structure as in the debate and philosophy cases. Deviations from the debate model are particularly notable for the r/news and r/science subreddits, where discussions appear to use the branching comment trees to structure semantic exploration. Meanwhile, r/The\_Donald shows more debate- and philosophy-like excursion patterns, particularly at the largest scales.}
  \label{reddit_table}
\end{table}

\clearpage
\section{Discussion}


To come together, exchange ideas, and argue is a natural part of human social life. These processes are defined not only by the ideas in play, but also the order in which they appear. Social institutions determine this order by attempting to reconcile differing preferences of participants with constraints of the environment.


A key result of this work is the characterization of how groups explore and exploit a space of ideas on different scales. Changing scales allows us to compare how ideas are explored in the sequential presentation of sentences, speeches, or comments as they are linked together in larger structures.

When we view discussions on larger scales we see greater order; the position of an element influences that of its successor more strongly at the paragraph level than at the sentence level. The emergence of order at larger scales differs from what one would expect from a purely spatial intuition that views discussion as ``walking'' over a space and slowly decorrelating in time, as in a simple diffusion model. In contrast, our empirical findings support the notion that it is possible to write down social grammars that govern these larger scales, such as those found in \citeA{jackendoff2007language}.

The presence of more explicit structure on larger scales---and its relative absence on small scales---may be due to the different ways in which individuals can process information. On the shortest timescales we hear sentences. Both the brevity of such units and the specialized modules of our brains that processes it may allow us to knit together more complex sequences into a meaningful whole, enabling greater exploration. On longer timescales, where the decoding of structure must be deliberative and learned through socialization, we need more guidance. If they are to be understood, discussants must explicitly connect their contributions to what has been said before, often using implicit signaling conventions that are invisible to our semantic models at the grammatical level. In this way, social grammars are less innate than lexical ones, and therefore subject to more cumbersome, extrinsically-imposed cognitive constraints.


As we go to larger scales, the $\mu$ and $\sigma$ parameters that describe group discussion consistently approach a limit region. In the language of physics, the change of scales is known as renormalization~\cite{huggett1995renormalisation,dedeo,DeDeo2018}; here, under renormalization, the parameters appear to ``flow'' to a small region in $\mu$ and $\sigma$ space, suggesting that the basic properties of a discussion are invariant to further changes in scale~\cite{wilson1975renormalization, kadanoffstatistical, caobook}. Moving from paragraph to paragraph, for example, has similar explore-exploit properties as moving from section-to-section or chapter to chapter.

The emergent order we detect connects to the idea of cognition on scales that transcend the individual. Our analysis looks at these systems from the point of view of a group-level mind~\cite{hutchins1995cognition, clark2010supersizing, theiner2011res, dedeo2014group}, finding structure without preserving the boundaries between individuals.

While we can imagine how an author might attempt to structure an argument in this way over the course of a book, it is more mysterious how it might happen in the distributed systems corresponding to debates and online discussion. In the case of the parliamentary debates, for example, the participants may find themselves on different political sides, wishing to wrest control of the discussion without regard for how this might confuse an observer. On the other hand, participants in these debates have no choice but to structure things: gaining control of the course of an argument may require the participant to follow what has come before. This fits with the results of \citeA{barron} and \citeA{elise}: what is new is (usually) quickly forgotten, and those who consume information show strong preferences for what looks like things they have already encountered.

The comparison between in-person debates and the online discussions is particularly illuminating. On the largest scales, samples from the political subreddit r/The\_Donald look very similar to those from the French parliament. Users appear to be neglecting the branching structure implicit in their choices of which comment to reply to, interacting with each other in a continuous stream. The r/science and r/news discussions show the lowest focus, and highest variance, of all the systems considered here. If, at the individual level, the participants are to process the information presented in the same way they process a philosophical text or an in-person debate, there must be additional guiding structure over and above simple temporal order.

Inspection validates the intuition. Exchanges on r/The\_Donald are often limited to the presentation of arguments and positions familiar to participants, with only the occasional clarifying question or criticism. Discussions on r/science and r/news, meanwhile, seem to use the comment structure in a self-organizing fashion that goes beyond the linear exchanges possible in a debate. Readers interested in a particular feature of the story in question join a relevant branch of the tree. Our evidence suggests they use these branches to engage in more detailed study, or debate, and can thus explore different parts of semantic space simultaneously. 

This greater reliance on branching structure can also be seen in the tree depth statistics themselves, Fig.~\ref{depth_dist}. All three of the subreddits have a non-trivial tree structure, and at least 50\% of the comments are nested at least one layer deep (\emph{i.e.}, are replies to other comments). The r/science and r/news comments, however, tend to nest deeper; while around 5\% of comments in these subreddits nest ten-layers deep, less than 1\% of r/The\_Donald do. Users on the former two systems find utility in maintaining separate branches for longer, and take them deeper.

Our results show that the ways in which users choose to nest their comments have semantic implications. Branching systems can segregate users in ways that frustrate a linear interpretation of thread development. Fig.~\ref{reddit_depth} shows that this relationship is not deterministic; the norms of discussion on each subreddit mean that even when comment patterns have similar depth distribution, a wide variation in discussion focus is possible. Depth only explains some of the variance in the development of discussions; including subreddit labels (\emph{i.e.}, considering each subreddit in turn) reduces the variance of the relationship futher. Topology matters, but so does the way the users co-construct and interact with it.

Why doesn't The\_Donald use the comment system to branch and develop discussion in the same way as r/science and r/news? One answer concerns the creation and maintenance of common knowledge~\cite{fagin2004reasoning}. Individuals come to r/science and r/news to learn about the world and share what they know. These activities can be successful even if all participants are not ``on the same page''; as long as a few people are able to gain consensus on the topic of a particular branch, learning and sharing can take place.

Political organization is different. There, common knowledge is required for large-scale, mutually acknowledged consensus that is often essential to effective action; it is so essential that the need for it can even drive the physical layout of halls where political debate takes place~\cite{chwe2013rational}. Agents need to know what everyone else thinks in order to correctly represent the goals of the movement they participate in, and to make sure their own voices are heard in that development of those goals. Joint action is only possible if everyone knows what the action is to be, and are at the same time aware that everyone knows that everyone knows. Branching systems work against this process, because individuals can isolate themselves into subgroups, while remaining ignorant of the activities of others. 

If common knowledge pushes systems to more temporally-ordered and linear interactions, it means that a branching style of online commenting will play a more limited role in politics than in the creation and sharing of knowledge. The political power of sites such as 4Chan~\cite{nagle2017kill} argues in favor of this hypothesis. 4Chan has even less structure than Reddit; 4Chan discussion boards are almost entirely linear, with comments following each other in a single temporally-defined order. 

By contrast, we expect branching systems to lead to new forms of knowledge production in cases where common knowledge is not required. Examples are increasingly common in the modern era~\cite{nielsen2011reinventing}. The Polymath Project~\cite{poly} and sites such as MathOverflow~\cite{mathover} provide salient examples of these new forms of software-enabled institutions.

\section{Conclusions}

An enormous literature has arisen to analyze the institutions we use to manage discussions and the sharing of ideas. Political scientists study parliament halls; legal scholars, the rules of the courtroom; economists, the market; rhetoricians, linguists, and philosophers, the written or verbal argument. In each case, the system under study accomplishes its goals in part by virtue of the way it processes information: parliaments take in data from constituents, stakeholders and experts; lawyers present arguments, evidence, and testimony within strict legal boundaries; buyers and sellers exchange information about needs and abilities through the price mechanism; writers and philosophers have traditionally been constrained by the linearity of the page to the serial argument.

Basic extensions of Arrow's theorem to group-level reasoning mean that any institution we create will be, at best, an imperfect mechanism for aggregating beliefs. It will work well only within limited domains defined by the questions in play and the preferences of the individuals for how to answer them~\cite{list2012theory}. New populations, and new types of questions, therefore drive the creation of new institutions~\cite{taylor2009secular,fukuyama2011origins,bellah2012axial}. To be effective, these must respect the basic cognitive limits of the individuals who participate in them.

Books, debates, and comment trees are very different ways to solve these organizational problems of sharing and discussion. A simple mathematical framework, based on an analogy to random walks in physical space, allows us to compare them on a level field. Order, in each case, appears on the largest scales: it is when one takes the ``God's eye view'' over the progress of an argument that the semantic relationships between steps become clear. This structure emerges from the more subtle, and more apparently chaotic, ways in which the same arguments unfold from sentence to sentence. This order is emergent, not inherent. It corresponds to a trans-linguistic order, driven by the technologies and social norms of communication, and the grammars that accompany them.

\section*{Acknowledgements}

We thank Gabe Salmon for helpful discussions at the beginning of this research and readings of this work in draft form. WHWT acknowledges the support of an Ariel Scholarship through St. John's College Santa Fe, and the Research Experience for Undergraduates program at the Santa Fe Institute under National Science Foundation Award \#ACI-1358567.

\clearpage 

\appendix
\section*{Appendix: Sentence, Speech, and Post Length Distributions}
\label{app_dist}

\begin{figure}[!h]
  \centering
  \begin{tabular}{cc}
    \includegraphics[width=0.45\linewidth]{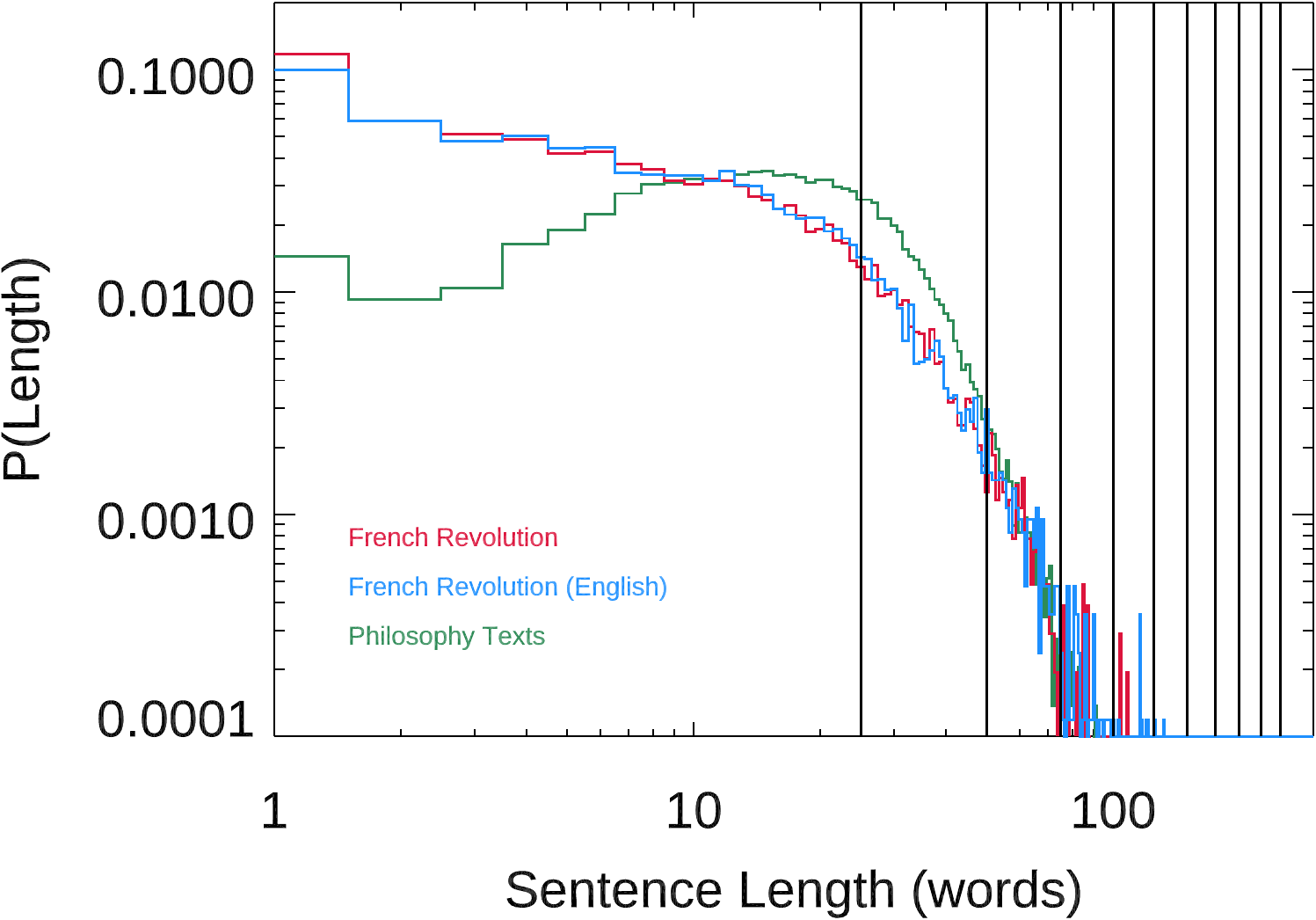}  &   \includegraphics[width=0.45\linewidth]{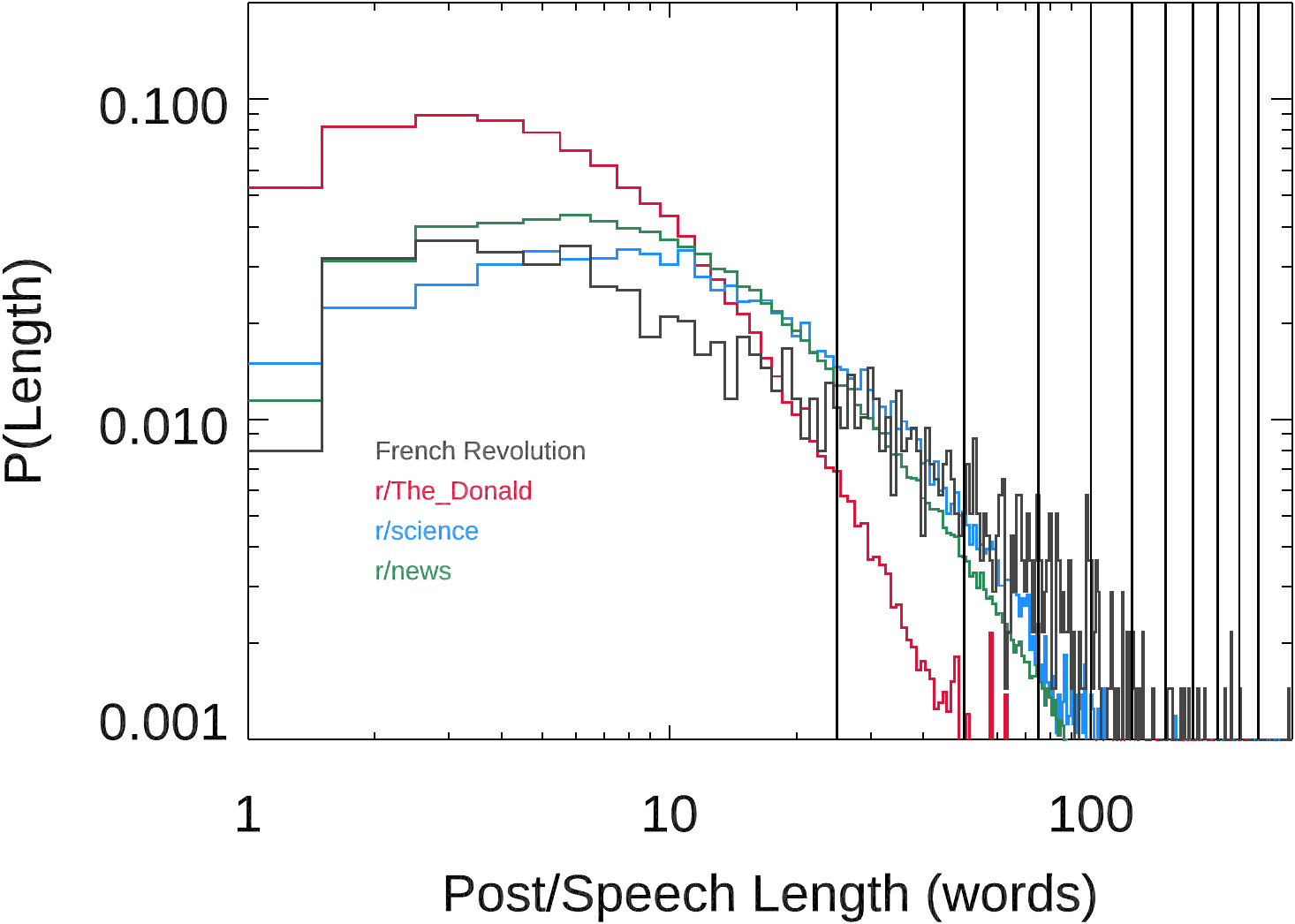} 
\end{tabular}
  \caption{{\bf Left}: The distribution of sentence lengths in the philosophical texts and French Revolution speeches. Overlaid are the chunk sizes used in this analysis. Above chunk size of 100, chunks always include at least one sentence. {\bf Right}: The distribution of speech and post lengths in the French Revolution speeches and the three subreddits under consideration. Revolutionary speeches have a length distribution comparable to posts on r/news and r/science, while r/The\_Donald favors shorter posts. Again, above chunk size of one hundred, chunks (almost) always include at least one post or speech.}
  \label{word_dist}
\end{figure}

\begin{figure}[!h]
  \centering
  \includegraphics[width=0.45\linewidth]{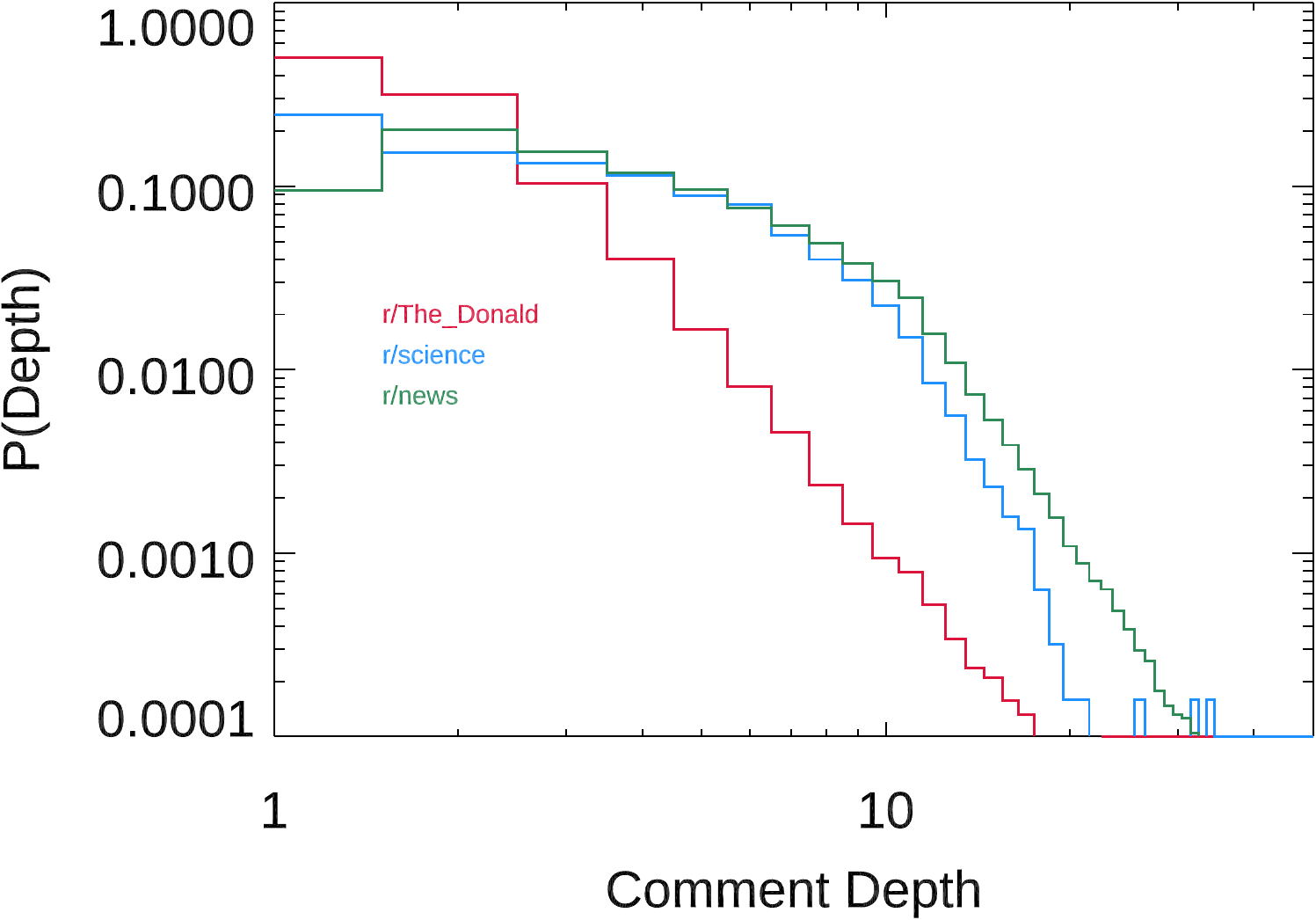}
  \caption{The distribution of comment depths in the three subreddits. A comment depth of one is attached to the submission itself; a comment with depth two is a reply to one of these comments, and so on. All three subreddits show significant levels of nesting and at least 50\% of all comments are replies to to others; r/science and r/news are particularly deep.}
  \label{depth_dist}
\end{figure}

\clearpage
\section*{Appendix: Bayesian Model-Fitting}
\label{app_bayes}

Our model is generative, which means we can estimate its two free parameters in a Bayesian fashion. In particular, we compute
\begin{equation}
\log{P(\mu, \sigma | \mathrm{data})} = \sum_{i=1}^{K-1} \log P(k_i | \mu, \sigma) + C,
\end{equation}
where the first term is the log-likelihood of the observed chunk-to-chunk surprises, $\{k_i\}$, given the distribution of surprises implied by the model, and the second term includes the constant offset from a flat prior over $\mu$ and $\sigma$. To compute the first, the distributions implied by the model are generated by simulation.

We then compute the minimum L2-loss estimates for $\mu$ and $\sigma$ by averaging over the posterior, \emph{i.e.}, we estimate
\begin{eqnarray*}
\hat{\mu} & = & \int \mu P(\mu,\sigma | \mathrm{data})~d\mu~d\sigma \\
\hat{\sigma} & = & \int \sigma P(\mu,\sigma | \mathrm{data})~d\mu~d\sigma,
\end{eqnarray*}
where we approximate these integrals by repeated sampling from the posterior. Standard deviations for each of the parameters are estimated in a similar fashion,
\begin{eqnarray*}
\hat{\sigma^2}_\mu & = & \int (\mu-\hat{\mu})^2 P(\mu,\sigma | \mathrm{data})~d\mu~d\sigma \\
\hat{\sigma^2}_\sigma & = & \int (\sigma-\hat{\sigma})^2 P(\mu,\sigma | \mathrm{data})~d\mu~d\sigma,
\end{eqnarray*}
again, by sampling from the posterior. Optimized C code, {\tt levytopic}, to estimate these parameters for an arbitrary sequence of chunks will be released on publication.




\clearpage 

\begin{thebibliography}{}

\bibitem [\protect \citeauthoryear {%
Aristotle%
}{%
Aristotle%
}{%
{\protect \APACyear {-322}}%
}]{%
aristotle}
\APACinsertmetastar {%
aristotle}%
\begin{APACrefauthors}%
Aristotle.%
\end{APACrefauthors}%
\unskip\
\newblock
\APACrefYear{-322}.
\newblock
\APACrefbtitle {Rhetoric} {Rhetoric}\ (E\BPBI M.~Cope\ \BBA {} J\BPBI
  E.~Sandys, \BEDS{}).
\newblock
\APACaddressPublisher{Cambridge, UK}{Cambridge University Press (2018 Edition)}.
\PrintBackRefs{\CurrentBib}

\bibitem [\protect \citeauthoryear {%
Barron%
, Huang%
, Spang%
\BCBL {}\ \BBA {} DeDeo%
}{%
Barron%
\ \protect \BOthers {.}}{%
{\protect \APACyear {2018}}%
}]{%
barron}
\APACinsertmetastar {%
barron}%
\begin{APACrefauthors}%
Barron, A\BPBI T\BPBI J.%
, Huang, J.%
, Spang, R\BPBI L.%
\BCBL {}\ \BBA {} DeDeo, S.%
\end{APACrefauthors}%
\unskip\
\newblock
\APACrefYearMonthDay{2018}{}{}.
\newblock
{\BBOQ}\APACrefatitle {Individuals, institutions, and innovation in the debates
  of the {French Revolution}} {Individuals, institutions, and innovation in the
  debates of the {French Revolution}}.{\BBCQ}
\newblock
\APACjournalVolNumPages{Proceedings of the National Academy of
  Sciences}{115}{18}{4607--4612}.
\newblock
\begin{APACrefDOI} \doi{10.1073/pnas.1717729115} \end{APACrefDOI}
\PrintBackRefs{\CurrentBib}

\bibitem [\protect \citeauthoryear {%
Bellah%
\ \BBA {} Joas%
}{%
Bellah%
\ \BBA {} Joas%
}{%
{\protect \APACyear {2012}}%
}]{%
bellah2012axial}
\APACinsertmetastar {%
bellah2012axial}%
\begin{APACrefauthors}%
Bellah, R.%
\BCBT {}\ \BBA {} Joas, H.%
\end{APACrefauthors}%
\unskip\
\newblock
\APACrefYear{2012}.
\newblock
\APACrefbtitle {The {Axial Age} and Its Consequences} {The {Axial Age} and its
  consequences}.
\newblock
\APACaddressPublisher{}{Harvard University Press}.
\PrintBackRefs{\CurrentBib}

\bibitem [\protect \citeauthoryear {%
Benhamou%
}{%
Benhamou%
}{%
{\protect \APACyear {2007}}%
}]{%
benhamou2007many}
\APACinsertmetastar {%
benhamou2007many}%
\begin{APACrefauthors}%
Benhamou, S.%
\end{APACrefauthors}%
\unskip\
\newblock
\APACrefYearMonthDay{2007}{}{}.
\newblock
{\BBOQ}\APACrefatitle {How many animals really do the {L}\'{e}vy walk?} {How
  many animals really do the {L}\'{e}vy walk?}{\BBCQ}
\newblock
\APACjournalVolNumPages{Ecology}{88}{8}{1962--1969}.
\PrintBackRefs{\CurrentBib}

\bibitem [\protect \citeauthoryear {%
Blei%
, Ng%
\BCBL {}\ \BBA {} Jordan%
}{%
Blei%
\ \protect \BOthers {.}}{%
{\protect \APACyear {2003}}%
}]{%
blei}
\APACinsertmetastar {%
blei}%
\begin{APACrefauthors}%
Blei, D\BPBI M.%
, Ng, A\BPBI Y.%
\BCBL {}\ \BBA {} Jordan, M\BPBI I.%
\end{APACrefauthors}%
\unskip\
\newblock
\APACrefYearMonthDay{2003}{}{}.
\newblock
{\BBOQ}\APACrefatitle {Latent dirichlet allocation} {Latent dirichlet
  allocation}.{\BBCQ}
\newblock
\APACjournalVolNumPages{Journal of Machine Learning Research}{3}{}{993--1022}.
\newblock
\APACrefnote{ACMid 944937}
\PrintBackRefs{\CurrentBib}

\bibitem [\protect \citeauthoryear {%
Chomsky%
}{%
Chomsky%
}{%
{\protect \APACyear {1965}}%
}]{%
chomsky2014aspects}
\APACinsertmetastar {%
chomsky2014aspects}%
\begin{APACrefauthors}%
Chomsky, N.%
\end{APACrefauthors}%
\unskip\
\newblock
\APACrefYear{1965}.
\newblock
\APACrefbtitle {Aspects of the Theory of Syntax} {Aspects of the theory of
  syntax}.
\newblock
\APACaddressPublisher{Cambridge, MA, USA}{MIT press}.
\newblock
\APACrefnote{2014 edition}
\PrintBackRefs{\CurrentBib}

\bibitem [\protect \citeauthoryear {%
Chwe%
}{%
Chwe%
}{%
{\protect \APACyear {2013}}%
}]{%
chwe2013rational}
\APACinsertmetastar {%
chwe2013rational}%
\begin{APACrefauthors}%
Chwe, M\BPBI S\BHBI Y.%
\end{APACrefauthors}%
\unskip\
\newblock
\APACrefYear{2013}.
\newblock
\APACrefbtitle {Rational ritual: Culture, coordination, and common knowledge}
  {Rational ritual: Culture, coordination, and common knowledge}.
\newblock
\APACaddressPublisher{Princeton, NJ, USA}{Princeton University Press}.
\PrintBackRefs{\CurrentBib}

\bibitem [\protect \citeauthoryear {%
Clark%
}{%
Clark%
}{%
{\protect \APACyear {2010}}%
}]{%
clark2010supersizing}
\APACinsertmetastar {%
clark2010supersizing}%
\begin{APACrefauthors}%
Clark, A.%
\end{APACrefauthors}%
\unskip\
\newblock
\APACrefYear{2010}.
\newblock
\APACrefbtitle {Supersizing the Mind: Embodiment, Action, and Cognitive
  Extension} {Supersizing the mind: Embodiment, action, and cognitive
  extension}.
\newblock
\APACaddressPublisher{}{Oxford University Press}.
\PrintBackRefs{\CurrentBib}

\bibitem [\protect \citeauthoryear {%
Cranshaw%
\ \BBA {} Kittur%
}{%
Cranshaw%
\ \BBA {} Kittur%
}{%
{\protect \APACyear {2011}}%
}]{%
poly}
\APACinsertmetastar {%
poly}%
\begin{APACrefauthors}%
Cranshaw, J.%
\BCBT {}\ \BBA {} Kittur, A.%
\end{APACrefauthors}%
\unskip\
\newblock
\APACrefYearMonthDay{2011}{}{}.
\newblock
{\BBOQ}\APACrefatitle {The {Polymath Project}: Lessons from a Successful Online
  Collaboration in Mathematics} {The {Polymath Project}: Lessons from a
  successful online collaboration in mathematics}.{\BBCQ}
\newblock
\BIn{} \APACrefbtitle {Proceedings of the {SIGCHI} Conference on Human Factors
  in Computing Systems} {Proceedings of the {SIGCHI} conference on human
  factors in computing systems}\ (\BPGS\ 1865--1874).
\newblock
\APACaddressPublisher{New York, NY, USA}{ACM}.
\newblock
\begin{APACrefDOI} \doi{10.1145/1978942.1979213} \end{APACrefDOI}
\PrintBackRefs{\CurrentBib}

\bibitem [\protect \citeauthoryear {%
DeDeo%
}{%
DeDeo%
}{%
{\protect \APACyear {2014}}%
}]{%
dedeo2014group}
\APACinsertmetastar {%
dedeo2014group}%
\begin{APACrefauthors}%
DeDeo, S.%
\end{APACrefauthors}%
\unskip\
\newblock
\APACrefYearMonthDay{2014}{}{}.
\newblock
{\BBOQ}\APACrefatitle {Group Minds and the Case of {W}ikipedia} {Group minds
  and the case of {W}ikipedia}.{\BBCQ}
\newblock
\APACjournalVolNumPages{Human Computation}{1}{1}{}.
\newblock
\APACrefnote{arXiv:1407.2210}
\newblock
\begin{APACrefDOI} \doi{10.15346/hc.v1i1.2} \end{APACrefDOI}
\PrintBackRefs{\CurrentBib}

\bibitem [\protect \citeauthoryear {%
DeDeo%
}{%
DeDeo%
}{%
{\protect \APACyear {2017}}%
}]{%
dedeo}
\APACinsertmetastar {%
dedeo}%
\begin{APACrefauthors}%
DeDeo, S.%
\end{APACrefauthors}%
\unskip\
\newblock
\APACrefYearMonthDay{2017}{}{}.
\newblock
{\BBOQ}\APACrefatitle {Major Transitions in Political Order} {Major transitions
  in political order}.{\BBCQ}
\newblock
\BIn{} S.~Walker, P.~Davies\BCBL {}\ \BBA {} G.~Ellis\ (\BEDS), \APACrefbtitle
  {From Matter to Life: Information and Causality.} {From matter to life:
  Information and causality.}
\newblock
\APACaddressPublisher{Cambridge, United Kingdom}{Cambridge University Press}.
\newblock
\APACrefnote{arXiv:1512.03419}
\newblock
\begin{APACrefDOI} \doi{10.1017/9781316584200.016} \end{APACrefDOI}
\PrintBackRefs{\CurrentBib}

\bibitem [\protect \citeauthoryear {%
DeDeo%
}{%
DeDeo%
}{%
{\protect \APACyear {2018}}%
}]{%
DeDeo2018}
\APACinsertmetastar {%
DeDeo2018}%
\begin{APACrefauthors}%
DeDeo, S.%
\end{APACrefauthors}%
\unskip\
\newblock
\APACrefYearMonthDay{2018}{}{}.
\newblock
{\BBOQ}\APACrefatitle {Origin Gaps and the Eternal Sunshine of the Second-Order
  Pendulum} {Origin gaps and the eternal sunshine of the second-order
  pendulum}.{\BBCQ}
\newblock
\BIn{} A.~Aguirre, B.~Foster\BCBL {}\ \BBA {} Z.~Merali\ (\BEDS),
  \APACrefbtitle {Wandering Towards a Goal: How Can Mindless Mathematical Laws
  Give Rise to Aims and Intention?} {Wandering towards a goal: How can mindless
  mathematical laws give rise to aims and intention?}\ (\BPGS\ 41--61).
\newblock
\APACaddressPublisher{Cham}{Springer International Publishing}.
\newblock
\begin{APACrefDOI} \doi{10.1007/978-3-319-75726-1\_5} \end{APACrefDOI}
\PrintBackRefs{\CurrentBib}

\bibitem [\protect \citeauthoryear {%
Fagin%
, Halpern%
, Moses%
\BCBL {}\ \BBA {} Vardi%
}{%
Fagin%
\ \protect \BOthers {.}}{%
{\protect \APACyear {2004}}%
}]{%
fagin2004reasoning}
\APACinsertmetastar {%
fagin2004reasoning}%
\begin{APACrefauthors}%
Fagin, R.%
, Halpern, J\BPBI Y.%
, Moses, Y.%
\BCBL {}\ \BBA {} Vardi, M.%
\end{APACrefauthors}%
\unskip\
\newblock
\APACrefYear{2004}.
\newblock
\APACrefbtitle {Reasoning about knowledge} {Reasoning about knowledge}.
\newblock
\APACaddressPublisher{Cambridge, MA, USA}{MIT Press}.
\PrintBackRefs{\CurrentBib}

\bibitem [\protect \citeauthoryear {%
Fisher%
}{%
Fisher%
}{%
{\protect \APACyear {2004}}%
}]{%
caobook}
\APACinsertmetastar {%
caobook}%
\begin{APACrefauthors}%
Fisher, M\BPBI E.%
\end{APACrefauthors}%
\unskip\
\newblock
\APACrefYearMonthDay{2004}{}{}.
\newblock
{\BBOQ}\APACrefatitle {Renormalization group theory: its basis and formulation
  in statistical physics} {Renormalization group theory: its basis and
  formulation in statistical physics}.{\BBCQ}
\newblock
\BIn{} T\BPBI Y.~Cao\ (\BED), \APACrefbtitle {Conceptual Foundations of Quantum
  Field Theory.} {Conceptual foundations of quantum field theory.}
\newblock
\APACaddressPublisher{Cambridge, UK}{Cambridge University Press}.
\PrintBackRefs{\CurrentBib}

\bibitem [\protect \citeauthoryear {%
Fukuyama%
}{%
Fukuyama%
}{%
{\protect \APACyear {2011}}%
}]{%
fukuyama2011origins}
\APACinsertmetastar {%
fukuyama2011origins}%
\begin{APACrefauthors}%
Fukuyama, F.%
\end{APACrefauthors}%
\unskip\
\newblock
\APACrefYear{2011}.
\newblock
\APACrefbtitle {{The Origins of Political Order: From Prehuman Times to the
  French Revolution}} {{The Origins of Political Order: From Prehuman Times to
  the French Revolution}}.
\newblock
\APACaddressPublisher{}{Farrar, Straus and Giroux}.
\PrintBackRefs{\CurrentBib}

\bibitem [\protect \citeauthoryear {%
Goldstone%
, Wisdom%
, Roberts%
\BCBL {}\ \BBA {} Frey%
}{%
Goldstone%
\ \protect \BOthers {.}}{%
{\protect \APACyear {2013}}%
}]{%
goldstone2013learning}
\APACinsertmetastar {%
goldstone2013learning}%
\begin{APACrefauthors}%
Goldstone, R\BPBI L.%
, Wisdom, T\BPBI N.%
, Roberts, M\BPBI E.%
\BCBL {}\ \BBA {} Frey, S.%
\end{APACrefauthors}%
\unskip\
\newblock
\APACrefYearMonthDay{2013}{}{}.
\newblock
{\BBOQ}\APACrefatitle {Learning along with others} {Learning along with
  others}.{\BBCQ}
\newblock
\BIn{} \APACrefbtitle {Psychology of Learning and Motivation} {Psychology of
  learning and motivation}\ (\BVOL~58, \BPGS\ 1--45).
\newblock
\APACaddressPublisher{}{Elsevier}.
\PrintBackRefs{\CurrentBib}

\bibitem [\protect \citeauthoryear {%
Grice%
}{%
Grice%
}{%
{\protect \APACyear {1975}}%
}]{%
grice1975logic}
\APACinsertmetastar {%
grice1975logic}%
\begin{APACrefauthors}%
Grice, H\BPBI P.%
\end{APACrefauthors}%
\unskip\
\newblock
\APACrefYearMonthDay{1975}{}{}.
\newblock
{\BBOQ}\APACrefatitle {Logic and conversation} {Logic and conversation}.{\BBCQ}
\newblock
\APACjournalVolNumPages{1975}{}{}{41--58}.
\PrintBackRefs{\CurrentBib}

\bibitem [\protect \citeauthoryear {%
Huggett%
\ \BBA {} Weingard%
}{%
Huggett%
\ \BBA {} Weingard%
}{%
{\protect \APACyear {1995}}%
}]{%
huggett1995renormalisation}
\APACinsertmetastar {%
huggett1995renormalisation}%
\begin{APACrefauthors}%
Huggett, N.%
\BCBT {}\ \BBA {} Weingard, R.%
\end{APACrefauthors}%
\unskip\
\newblock
\APACrefYearMonthDay{1995}{}{}.
\newblock
{\BBOQ}\APACrefatitle {The renormalisation group and effective field theories}
  {The renormalisation group and effective field theories}.{\BBCQ}
\newblock
\APACjournalVolNumPages{Synthese}{102}{1}{171--194}.
\PrintBackRefs{\CurrentBib}

\bibitem [\protect \citeauthoryear {%
Hutchins%
\ \BBA {} Press%
}{%
Hutchins%
\ \BBA {} Press%
}{%
{\protect \APACyear {1995}}%
}]{%
hutchins1995cognition}
\APACinsertmetastar {%
hutchins1995cognition}%
\begin{APACrefauthors}%
Hutchins, E.%
\BCBT {}\ \BBA {} Press, M.%
\end{APACrefauthors}%
\unskip\
\newblock
\APACrefYear{1995}.
\newblock
\APACrefbtitle {Cognition in the Wild} {Cognition in the wild}.
\newblock
\APACaddressPublisher{}{MIT Press}.
\PrintBackRefs{\CurrentBib}

\bibitem [\protect \citeauthoryear {%
Itti%
\ \BBA {} Baldi%
}{%
Itti%
\ \BBA {} Baldi%
}{%
{\protect \APACyear {2009}}%
}]{%
itti2009bayesian}
\APACinsertmetastar {%
itti2009bayesian}%
\begin{APACrefauthors}%
Itti, L.%
\BCBT {}\ \BBA {} Baldi, P.%
\end{APACrefauthors}%
\unskip\
\newblock
\APACrefYearMonthDay{2009}{}{}.
\newblock
{\BBOQ}\APACrefatitle {Bayesian surprise attracts human attention} {Bayesian
  surprise attracts human attention}.{\BBCQ}
\newblock
\APACjournalVolNumPages{Vision research}{49}{10}{1295--1306}.
\PrintBackRefs{\CurrentBib}

\bibitem [\protect \citeauthoryear {%
Jackendoff%
}{%
Jackendoff%
}{%
{\protect \APACyear {2002}}%
}]{%
jackendoff2002foundations}
\APACinsertmetastar {%
jackendoff2002foundations}%
\begin{APACrefauthors}%
Jackendoff, R.%
\end{APACrefauthors}%
\unskip\
\newblock
\APACrefYear{2002}.
\newblock
\APACrefbtitle {Foundations of Language: Brain, Meaning, Grammar, Evolution}
  {Foundations of language: Brain, meaning, grammar, evolution}.
\newblock
\APACaddressPublisher{Oxford, UK}{Oxford University Press}.
\PrintBackRefs{\CurrentBib}

\bibitem [\protect \citeauthoryear {%
Jackendoff%
}{%
Jackendoff%
}{%
{\protect \APACyear {2007}}%
}]{%
jackendoff2007language}
\APACinsertmetastar {%
jackendoff2007language}%
\begin{APACrefauthors}%
Jackendoff, R.%
\end{APACrefauthors}%
\unskip\
\newblock
\APACrefYear{2007}.
\newblock
\APACrefbtitle {Language, consciousness, culture: Essays on mental structure}
  {Language, consciousness, culture: Essays on mental structure}.
\newblock
\APACaddressPublisher{Cambridge, MA, USA}{MIT Press}.
\PrintBackRefs{\CurrentBib}

\bibitem [\protect \citeauthoryear {%
Jaeger%
}{%
Jaeger%
}{%
{\protect \APACyear {2010}}%
}]{%
wedel2}
\APACinsertmetastar {%
wedel2}%
\begin{APACrefauthors}%
Jaeger, T\BPBI F.%
\end{APACrefauthors}%
\unskip\
\newblock
\APACrefYearMonthDay{2010}{}{}.
\newblock
{\BBOQ}\APACrefatitle {Redundancy and reduction: Speakers manage syntactic
  information density} {Redundancy and reduction: Speakers manage syntactic
  information density}.{\BBCQ}
\newblock
\APACjournalVolNumPages{Cognitive Psychology}{61}{1}{23--62}.
\PrintBackRefs{\CurrentBib}

\bibitem [\protect \citeauthoryear {%
Jaeger%
\ \BBA {} Buz%
}{%
Jaeger%
\ \BBA {} Buz%
}{%
{\protect \APACyear {2017}}%
}]{%
wedel3}
\APACinsertmetastar {%
wedel3}%
\begin{APACrefauthors}%
Jaeger, T\BPBI F.%
\BCBT {}\ \BBA {} Buz, E.%
\end{APACrefauthors}%
\unskip\
\newblock
\APACrefYearMonthDay{2017}{}{}.
\newblock
{\BBOQ}\APACrefatitle {Signal reduction and linguistic encoding} {Signal
  reduction and linguistic encoding}.{\BBCQ}
\newblock
\BIn{} E.~Fern{\'a}ndez\ \BBA {} H.~Cairns\ (\BEDS), \APACrefbtitle {The
  Handbook of Psycholinguistics.} {The handbook of psycholinguistics.}
\newblock
\APACaddressPublisher{Hoboken, NJ, USA}{John Wiley \& Sons}.
\PrintBackRefs{\CurrentBib}

\bibitem [\protect \citeauthoryear {%
Jaeger%
\ \BBA {} Levy%
}{%
Jaeger%
\ \BBA {} Levy%
}{%
{\protect \APACyear {2007}}%
}]{%
wedel5}
\APACinsertmetastar {%
wedel5}%
\begin{APACrefauthors}%
Jaeger, T\BPBI F.%
\BCBT {}\ \BBA {} Levy, R\BPBI P.%
\end{APACrefauthors}%
\unskip\
\newblock
\APACrefYearMonthDay{2007}{}{}.
\newblock
{\BBOQ}\APACrefatitle {Speakers optimize information density through syntactic
  reduction} {Speakers optimize information density through syntactic
  reduction}.{\BBCQ}
\newblock
\BIn{} B.~Sch\"{o}lkopf, J\BPBI C.~Platt\BCBL {}\ \BBA {} T.~Hoffman\ (\BEDS),
  \APACrefbtitle {Advances in Neural Information Processing Systems 19}
  {Advances in neural information processing systems 19}\ (\BPGS\ 849--856).
\newblock
\APACaddressPublisher{}{MIT Press}.
\PrintBackRefs{\CurrentBib}

\bibitem [\protect \citeauthoryear {%
Jing%
, DeDeo%
\BCBL {}\ \BBA {} Ahn%
}{%
Jing%
\ \protect \BOthers {.}}{%
{\protect \APACyear {2018}}%
}]{%
elise}
\APACinsertmetastar {%
elise}%
\begin{APACrefauthors}%
Jing, E.%
, DeDeo, S.%
\BCBL {}\ \BBA {} Ahn, Y\BHBI Y.%
\end{APACrefauthors}%
\unskip\
\newblock
\APACrefYearMonthDay{2018}{}{}.
\newblock
{\BBOQ}\APACrefatitle {Sameness Attracts, Novelty Disturbs, but Outliers
  Flourish in Fanfiction Online} {Sameness attracts, novelty disturbs, but
  outliers flourish in fanfiction online}.{\BBCQ}
\newblock
\APACjournalVolNumPages{International AAAI Conference on Web and Social
  Media}{}{}{}.
\newblock
\APACrefnote{In review}
\PrintBackRefs{\CurrentBib}

\bibitem [\protect \citeauthoryear {%
Jurafsky%
}{%
Jurafsky%
}{%
{\protect \APACyear {1996}}%
}]{%
wedel4}
\APACinsertmetastar {%
wedel4}%
\begin{APACrefauthors}%
Jurafsky, D.%
\end{APACrefauthors}%
\unskip\
\newblock
\APACrefYearMonthDay{1996}{}{}.
\newblock
{\BBOQ}\APACrefatitle {A probabilistic model of lexical and syntactic access
  and disambiguation} {A probabilistic model of lexical and syntactic access
  and disambiguation}.{\BBCQ}
\newblock
\APACjournalVolNumPages{Cognitive Science}{20}{2}{137--194}.
\PrintBackRefs{\CurrentBib}

\bibitem [\protect \citeauthoryear {%
Kadanoff%
}{%
Kadanoff%
}{%
{\protect \APACyear {2000}}%
}]{%
kadanoffstatistical}
\APACinsertmetastar {%
kadanoffstatistical}%
\begin{APACrefauthors}%
Kadanoff, L\BPBI P.%
\end{APACrefauthors}%
\unskip\
\newblock
\APACrefYear{2000}.
\newblock
\APACrefbtitle {Statistical Physics: statics, dynamics and renormalization}
  {Statistical physics: statics, dynamics and renormalization}.
\newblock
\APACaddressPublisher{}{World Scientific}.
\PrintBackRefs{\CurrentBib}

\bibitem [\protect \citeauthoryear {%
Kleinberg%
}{%
Kleinberg%
}{%
{\protect \APACyear {2000}}%
}]{%
kleinberg2000navigation}
\APACinsertmetastar {%
kleinberg2000navigation}%
\begin{APACrefauthors}%
Kleinberg, J\BPBI M.%
\end{APACrefauthors}%
\unskip\
\newblock
\APACrefYearMonthDay{2000}{}{}.
\newblock
{\BBOQ}\APACrefatitle {Navigation in a small world} {Navigation in a small
  world}.{\BBCQ}
\newblock
\APACjournalVolNumPages{Nature}{406}{6798}{845}.
\PrintBackRefs{\CurrentBib}

\bibitem [\protect \citeauthoryear {%
Lanham%
}{%
Lanham%
}{%
{\protect \APACyear {2012}}%
}]{%
lanham2012handlist}
\APACinsertmetastar {%
lanham2012handlist}%
\begin{APACrefauthors}%
Lanham, R.%
\end{APACrefauthors}%
\unskip\
\newblock
\APACrefYear{2012}.
\newblock
\APACrefbtitle {A Handlist of Rhetorical Terms} {A handlist of rhetorical
  terms}.
\newblock
\APACaddressPublisher{Berkeley, CA, USA}{University of California Press}.
\PrintBackRefs{\CurrentBib}

\bibitem [\protect \citeauthoryear {%
List%
}{%
List%
}{%
{\protect \APACyear {2012}}%
}]{%
list2012theory}
\APACinsertmetastar {%
list2012theory}%
\begin{APACrefauthors}%
List, C.%
\end{APACrefauthors}%
\unskip\
\newblock
\APACrefYearMonthDay{2012}{}{}.
\newblock
{\BBOQ}\APACrefatitle {The theory of judgment aggregation: An introductory
  review} {The theory of judgment aggregation: An introductory review}.{\BBCQ}
\newblock
\APACjournalVolNumPages{Synthese}{187}{1}{179--207}.
\PrintBackRefs{\CurrentBib}

\bibitem [\protect \citeauthoryear {%
Lorince%
\ \BBA {} Todd%
}{%
Lorince%
\ \BBA {} Todd%
}{%
{\protect \APACyear {2016}}%
}]{%
lorince2016music}
\APACinsertmetastar {%
lorince2016music}%
\begin{APACrefauthors}%
Lorince, J.%
\BCBT {}\ \BBA {} Todd, P.%
\end{APACrefauthors}%
\unskip\
\newblock
\APACrefYearMonthDay{2016}{}{}.
\newblock
{\BBOQ}\APACrefatitle {Music tagging and listening: Testing the memory cue
  hypothesis in a collaborative tagging system} {Music tagging and listening:
  Testing the memory cue hypothesis in a collaborative tagging system}.{\BBCQ}
\newblock
\BIn{} \APACrefbtitle {Big Data in Cognitive Science: From Methods to
  Insights.} {Big data in cognitive science: From methods to insights.}
\newblock
\APACaddressPublisher{}{Taylor \& Francis}.
\PrintBackRefs{\CurrentBib}

\bibitem [\protect \citeauthoryear {%
Murdock%
, Allen%
\BCBL {}\ \BBA {} DeDeo%
}{%
Murdock%
\ \protect \BOthers {.}}{%
{\protect \APACyear {2017}}%
}]{%
murdock2017exploration}
\APACinsertmetastar {%
murdock2017exploration}%
\begin{APACrefauthors}%
Murdock, J.%
, Allen, C.%
\BCBL {}\ \BBA {} DeDeo, S.%
\end{APACrefauthors}%
\unskip\
\newblock
\APACrefYearMonthDay{2017}{}{}.
\newblock
{\BBOQ}\APACrefatitle {{Exploration and exploitation of Victorian science in
  Darwin's reading notebooks}} {{Exploration and exploitation of Victorian
  science in Darwin's reading notebooks}}.{\BBCQ}
\newblock
\APACjournalVolNumPages{Cognition}{159}{}{117--126}.
\PrintBackRefs{\CurrentBib}

\bibitem [\protect \citeauthoryear {%
Murdock%
, Allen%
\BCBL {}\ \BBA {} DeDeo%
}{%
Murdock%
\ \protect \BOthers {.}}{%
{\protect \APACyear {2018}}%
}]{%
joint_model}
\APACinsertmetastar {%
joint_model}%
\begin{APACrefauthors}%
Murdock, J.%
, Allen, C.%
\BCBL {}\ \BBA {} DeDeo, S.%
\end{APACrefauthors}%
\unskip\
\newblock
\APACrefYearMonthDay{2018}{}{}.
\newblock
{\BBOQ}\APACrefatitle {{The Development of Darwin's Origin of Species}} {{The
  Development of Darwin's Origin of Species}}.{\BBCQ}
\newblock
\APACjournalVolNumPages{CoRR}{abs/1802.09944}{}{}.
\PrintBackRefs{\CurrentBib}

\bibitem [\protect \citeauthoryear {%
Nagle%
}{%
Nagle%
}{%
{\protect \APACyear {2017}}%
}]{%
nagle2017kill}
\APACinsertmetastar {%
nagle2017kill}%
\begin{APACrefauthors}%
Nagle, A.%
\end{APACrefauthors}%
\unskip\
\newblock
\APACrefYear{2017}.
\newblock
\APACrefbtitle {{Kill all normies: Online culture wars from 4chan and Tumblr to
  Trump and the alt-right}} {{Kill all normies: Online culture wars from 4chan
  and Tumblr to Trump and the alt-right}}.
\newblock
\APACaddressPublisher{New York, NY, USA}{John Hunt Publishing}.
\PrintBackRefs{\CurrentBib}

\bibitem [\protect \citeauthoryear {%
Nassar%
, Wilson%
, Heasly%
\BCBL {}\ \BBA {} Gold%
}{%
Nassar%
\ \protect \BOthers {.}}{%
{\protect \APACyear {2010}}%
}]{%
nassar2010approximately}
\APACinsertmetastar {%
nassar2010approximately}%
\begin{APACrefauthors}%
Nassar, M\BPBI R.%
, Wilson, R\BPBI C.%
, Heasly, B.%
\BCBL {}\ \BBA {} Gold, J\BPBI I.%
\end{APACrefauthors}%
\unskip\
\newblock
\APACrefYearMonthDay{2010}{}{}.
\newblock
{\BBOQ}\APACrefatitle {An approximately {B}ayesian delta-rule model explains
  the dynamics of belief updating in a changing environment} {An approximately
  {B}ayesian delta-rule model explains the dynamics of belief updating in a
  changing environment}.{\BBCQ}
\newblock
\APACjournalVolNumPages{{Journal of Neuroscience}}{30}{37}{12366--12378}.
\PrintBackRefs{\CurrentBib}

\bibitem [\protect \citeauthoryear {%
Nielsen%
}{%
Nielsen%
}{%
{\protect \APACyear {2011}}%
}]{%
nielsen2011reinventing}
\APACinsertmetastar {%
nielsen2011reinventing}%
\begin{APACrefauthors}%
Nielsen, M.%
\end{APACrefauthors}%
\unskip\
\newblock
\APACrefYear{2011}.
\newblock
\APACrefbtitle {Reinventing discovery: the new era of networked science}
  {Reinventing discovery: the new era of networked science}.
\newblock
\APACaddressPublisher{Princeton, NJ, USA}{Princeton University Press}.
\PrintBackRefs{\CurrentBib}

\bibitem [\protect \citeauthoryear {%
Pirolli%
}{%
Pirolli%
}{%
{\protect \APACyear {2007}}%
}]{%
pirolli2007information}
\APACinsertmetastar {%
pirolli2007information}%
\begin{APACrefauthors}%
Pirolli, P.%
\end{APACrefauthors}%
\unskip\
\newblock
\APACrefYear{2007}.
\newblock
\APACrefbtitle {Information Foraging Theory: Adaptive Interaction with
  Information} {Information foraging theory: Adaptive interaction with
  information}.
\newblock
\APACaddressPublisher{Oxford, UK}{Oxford University Press}.
\PrintBackRefs{\CurrentBib}

\bibitem [\protect \citeauthoryear {%
Reynolds%
\ \BBA {} Rhodes%
}{%
Reynolds%
\ \BBA {} Rhodes%
}{%
{\protect \APACyear {2009}}%
}]{%
reynolds2009levy}
\APACinsertmetastar {%
reynolds2009levy}%
\begin{APACrefauthors}%
Reynolds, A\BPBI M.%
\BCBT {}\ \BBA {} Rhodes, C\BPBI J.%
\end{APACrefauthors}%
\unskip\
\newblock
\APACrefYearMonthDay{2009}{}{}.
\newblock
{\BBOQ}\APACrefatitle {The {L}{\'e}vy flight paradigm: random search patterns
  and mechanisms} {The {L}{\'e}vy flight paradigm: random search patterns and
  mechanisms}.{\BBCQ}
\newblock
\APACjournalVolNumPages{Ecology}{90}{4}{877--887}.
\PrintBackRefs{\CurrentBib}

\bibitem [\protect \citeauthoryear {%
Sang%
, Todd%
, Goldstone%
\BCBL {}\ \BBA {} Hills%
}{%
Sang%
\ \protect \BOthers {.}}{%
{\protect \APACyear {2018}}%
}]{%
sang_todd_goldstone_hills_2018}
\APACinsertmetastar {%
sang_todd_goldstone_hills_2018}%
\begin{APACrefauthors}%
Sang, K.%
, Todd, P\BPBI M.%
, Goldstone, R.%
\BCBL {}\ \BBA {} Hills, T\BPBI T.%
\end{APACrefauthors}%
\unskip\
\newblock
\APACrefYearMonthDay{2018}{}{}.
\newblock
\APACrefbtitle {Explore/exploit tradeoff strategies in a resource accumulation
  search task.} {Explore/exploit tradeoff strategies in a resource accumulation
  search task.}
\newblock
\APACaddressPublisher{}{PsyArXiv}.
\newblock
\begin{APACrefDOI} \doi{10.31234/osf.io/zw3s8} \end{APACrefDOI}
\PrintBackRefs{\CurrentBib}

\bibitem [\protect \citeauthoryear {%
Searle%
}{%
Searle%
}{%
{\protect \APACyear {2010}}%
}]{%
searle2010making}
\APACinsertmetastar {%
searle2010making}%
\begin{APACrefauthors}%
Searle, J.%
\end{APACrefauthors}%
\unskip\
\newblock
\APACrefYear{2010}.
\newblock
\APACrefbtitle {Making the social world: The structure of human civilization}
  {Making the social world: The structure of human civilization}.
\newblock
\APACaddressPublisher{Oxford, UK}{Oxford University Press}.
\PrintBackRefs{\CurrentBib}

\bibitem [\protect \citeauthoryear {%
Shlesinger%
}{%
Shlesinger%
}{%
{\protect \APACyear {2006}}%
}]{%
shlesinger2006random}
\APACinsertmetastar {%
shlesinger2006random}%
\begin{APACrefauthors}%
Shlesinger, M\BPBI F.%
\end{APACrefauthors}%
\unskip\
\newblock
\APACrefYearMonthDay{2006}{}{}.
\newblock
{\BBOQ}\APACrefatitle {Random walks: follow the money} {Random walks: follow
  the money}.{\BBCQ}
\newblock
\APACjournalVolNumPages{Nature Physics}{2}{2}{69}.
\PrintBackRefs{\CurrentBib}

\bibitem [\protect \citeauthoryear {%
Sloman%
}{%
Sloman%
}{%
{\protect \APACyear {1996}}%
}]{%
sloman1996empirical}
\APACinsertmetastar {%
sloman1996empirical}%
\begin{APACrefauthors}%
Sloman, S\BPBI A.%
\end{APACrefauthors}%
\unskip\
\newblock
\APACrefYearMonthDay{1996}{}{}.
\newblock
{\BBOQ}\APACrefatitle {The empirical case for two systems of reasoning} {The
  empirical case for two systems of reasoning}.{\BBCQ}
\newblock
\APACjournalVolNumPages{{Psychological Bulletin}}{119}{1}{3}.
\newblock
\begin{APACrefDOI} \doi{10.1037/0033-2909.119.1.3} \end{APACrefDOI}
\PrintBackRefs{\CurrentBib}

\bibitem [\protect \citeauthoryear {%
Tackett%
}{%
Tackett%
}{%
{\protect \APACyear {2015}}%
}]{%
tackett2015coming}
\APACinsertmetastar {%
tackett2015coming}%
\begin{APACrefauthors}%
Tackett, T.%
\end{APACrefauthors}%
\unskip\
\newblock
\APACrefYear{2015}.
\newblock
\APACrefbtitle {{The coming of the Terror in the French Revolution}} {{The
  coming of the Terror in the French Revolution}}.
\newblock
\APACaddressPublisher{Cambridge, MA, USA}{Harvard University Press}.
\PrintBackRefs{\CurrentBib}

\bibitem [\protect \citeauthoryear {%
Tausczik%
, Kittur%
\BCBL {}\ \BBA {} Kraut%
}{%
Tausczik%
\ \protect \BOthers {.}}{%
{\protect \APACyear {2014}}%
}]{%
mathover}
\APACinsertmetastar {%
mathover}%
\begin{APACrefauthors}%
Tausczik, Y\BPBI R.%
, Kittur, A.%
\BCBL {}\ \BBA {} Kraut, R\BPBI E.%
\end{APACrefauthors}%
\unskip\
\newblock
\APACrefYearMonthDay{2014}{}{}.
\newblock
{\BBOQ}\APACrefatitle {Collaborative Problem Solving: A Study of
  {MathOverflow}} {Collaborative problem solving: A study of
  {MathOverflow}}.{\BBCQ}
\newblock
\BIn{} \APACrefbtitle {{Proceedings of the 17th ACM Conference on Computer
  Supported Cooperative Work \&\#38; Social Computing}} {{Proceedings of the
  17th ACM Conference on Computer Supported Cooperative Work \&\#38; Social
  Computing}}\ (\BPGS\ 355--367).
\newblock
\APACaddressPublisher{New York, NY, USA}{ACM}.
\newblock
\begin{APACrefDOI} \doi{10.1145/2531602.2531690} \end{APACrefDOI}
\PrintBackRefs{\CurrentBib}

\bibitem [\protect \citeauthoryear {%
Taylor%
}{%
Taylor%
}{%
{\protect \APACyear {2009}}%
}]{%
taylor2009secular}
\APACinsertmetastar {%
taylor2009secular}%
\begin{APACrefauthors}%
Taylor, C.%
\end{APACrefauthors}%
\unskip\
\newblock
\APACrefYear{2009}.
\newblock
\APACrefbtitle {A Secular Age} {A secular age}.
\newblock
\APACaddressPublisher{Cambridge, MA, USA}{Harvard University Press}.
\PrintBackRefs{\CurrentBib}

\bibitem [\protect \citeauthoryear {%
Theiner%
}{%
Theiner%
}{%
{\protect \APACyear {2011}}%
}]{%
theiner2011res}
\APACinsertmetastar {%
theiner2011res}%
\begin{APACrefauthors}%
Theiner, G.%
\end{APACrefauthors}%
\unskip\
\newblock
\APACrefYear{2011}.
\newblock
\APACrefbtitle {{Res Cogitans Extensa: A Philosophical Defense of the Extended
  Mind Thesis}} {{Res Cogitans Extensa: A Philosophical Defense of the Extended
  Mind Thesis}}.
\newblock
\APACaddressPublisher{}{Peter Lang, Internationaler Verlag der Wissenschaften}.
\PrintBackRefs{\CurrentBib}

\bibitem [\protect \citeauthoryear {%
Tomasello%
}{%
Tomasello%
}{%
{\protect \APACyear {2009}}%
}]{%
Tomasello}
\APACinsertmetastar {%
Tomasello}%
\begin{APACrefauthors}%
Tomasello, M.%
\end{APACrefauthors}%
\unskip\
\newblock
\APACrefYear{2009}.
\newblock
\APACrefbtitle {Why We Cooperate} {Why we cooperate}.
\newblock
\APACaddressPublisher{Cambridge, MA, USA}{MIT Press}.
\newblock
\APACrefnote{With responses by Carol Dweck, Joan Silk, Brian Skyrms, and
  Elizabeth Spelke.}
\PrintBackRefs{\CurrentBib}

\bibitem [\protect \citeauthoryear {%
Viswanathan%
\ \protect \BOthers {.}}{%
Viswanathan%
\ \protect \BOthers {.}}{%
{\protect \APACyear {1999}}%
}]{%
viswanathan1999optimizing}
\APACinsertmetastar {%
viswanathan1999optimizing}%
\begin{APACrefauthors}%
Viswanathan, G\BPBI M.%
, Buldyrev, S\BPBI V.%
, Havlin, S.%
, Da~Luz, M.%
, Raposo, E.%
\BCBL {}\ \BBA {} Stanley, H\BPBI E.%
\end{APACrefauthors}%
\unskip\
\newblock
\APACrefYearMonthDay{1999}{}{}.
\newblock
{\BBOQ}\APACrefatitle {Optimizing the success of random searches} {Optimizing
  the success of random searches}.{\BBCQ}
\newblock
\APACjournalVolNumPages{Nature}{401}{6756}{911}.
\PrintBackRefs{\CurrentBib}

\bibitem [\protect \citeauthoryear {%
Wedel%
, Jackson%
\BCBL {}\ \BBA {} Kaplan%
}{%
Wedel%
\ \protect \BOthers {.}}{%
{\protect \APACyear {2013}}%
}]{%
wedel1}
\APACinsertmetastar {%
wedel1}%
\begin{APACrefauthors}%
Wedel, A.%
, Jackson, S.%
\BCBL {}\ \BBA {} Kaplan, A.%
\end{APACrefauthors}%
\unskip\
\newblock
\APACrefYearMonthDay{2013}{}{}.
\newblock
{\BBOQ}\APACrefatitle {Functional load and the lexicon: Evidence that syntactic
  category and frequency relationships in minimal lemma pairs predict the loss
  of phoneme contrasts in language change} {Functional load and the lexicon:
  Evidence that syntactic category and frequency relationships in minimal lemma
  pairs predict the loss of phoneme contrasts in language change}.{\BBCQ}
\newblock
\APACjournalVolNumPages{Language and Speech}{56}{3}{395--417}.
\PrintBackRefs{\CurrentBib}

\bibitem [\protect \citeauthoryear {%
Wilson%
}{%
Wilson%
}{%
{\protect \APACyear {1975}}%
}]{%
wilson1975renormalization}
\APACinsertmetastar {%
wilson1975renormalization}%
\begin{APACrefauthors}%
Wilson, K\BPBI G.%
\end{APACrefauthors}%
\unskip\
\newblock
\APACrefYearMonthDay{1975}{}{}.
\newblock
{\BBOQ}\APACrefatitle {The {Renormalization Group}: Critical phenomena and the
  {K}ondo problem} {The {Renormalization Group}: Critical phenomena and the
  {K}ondo problem}.{\BBCQ}
\newblock
\APACjournalVolNumPages{{Reviews of Modern Physics}}{47}{4}{773}.
\newblock
\begin{APACrefDOI} \doi{10.1103/RevModPhys.47.773} \end{APACrefDOI}
\PrintBackRefs{\CurrentBib}

\bibitem [\protect \citeauthoryear {%
Wisdom%
, Song%
\BCBL {}\ \BBA {} Goldstone%
}{%
Wisdom%
\ \protect \BOthers {.}}{%
{\protect \APACyear {2013}}%
}]{%
wisdom2013social}
\APACinsertmetastar {%
wisdom2013social}%
\begin{APACrefauthors}%
Wisdom, T\BPBI N.%
, Song, X.%
\BCBL {}\ \BBA {} Goldstone, R\BPBI L.%
\end{APACrefauthors}%
\unskip\
\newblock
\APACrefYearMonthDay{2013}{}{}.
\newblock
{\BBOQ}\APACrefatitle {Social learning strategies in networked groups} {Social
  learning strategies in networked groups}.{\BBCQ}
\newblock
\APACjournalVolNumPages{Cognitive Science}{37}{8}{1383--1425}.
\PrintBackRefs{\CurrentBib}

\bibitem [\protect \citeauthoryear {%
Youn%
, Strumsky%
, Bettencourt%
\BCBL {}\ \BBA {} Lobo%
}{%
Youn%
\ \protect \BOthers {.}}{%
{\protect \APACyear {2015}}%
}]{%
youn2015invention}
\APACinsertmetastar {%
youn2015invention}%
\begin{APACrefauthors}%
Youn, H.%
, Strumsky, D.%
, Bettencourt, L\BPBI M.%
\BCBL {}\ \BBA {} Lobo, J.%
\end{APACrefauthors}%
\unskip\
\newblock
\APACrefYearMonthDay{2015}{}{}.
\newblock
{\BBOQ}\APACrefatitle {{Invention as a combinatorial process: evidence from US
  patents}} {{Invention as a combinatorial process: evidence from US
  patents}}.{\BBCQ}
\newblock
\APACjournalVolNumPages{Journal of The Royal Society
  Interface}{12}{106}{20150272}.
\PrintBackRefs{\CurrentBib}

\end{thebibliography}







\end{document}